%% file: SN2011ht.tex
\documentclass[12pt,preprint]{aastex}

\shorttitle{SN2011ht}
\shortauthors{R. Humphreys et al. }

\begin{document}

\title{The Unusual Temporal and Spectral Evolution of SN2011ht. II. Peculiar Type IIn or Impostor?  \altaffilmark{1}}

\author{
Roberta M. Humphreys,\altaffilmark{2}
Kris Davidson, \altaffilmark{2}
Terry J. Jones, \altaffilmark{2}
R. W. Pogge,  \altaffilmark{3}
Skyler H. Grammer, \altaffilmark{2}
Jos\'e L. Prieto,\altaffilmark{4}
T. A. Pritchard, \altaffilmark{5}
}

\altaffiltext{1}  
{Based  on observations  with the Large Binocular Telescope (LBT), an international collaboration among institutions in the United
States, Italy and Germany. LBT Corporation partners are: The University of
Arizona on behalf of the Arizona university system; Istituto Nazionale di
Astrofisica, Italy; LBT Beteiligungsgesellschaft, Germany, representing the
Max-Planck Society, the Astrophysical Institute Potsdam, and Heidelberg
University; The Ohio State University, and The Research Corporation, on
behalf of The University of Notre Dame, University of Minnesota and
University of Virginia. Based also on observations obtained at the MMT Observatory, a joint
facility of the Smithsonian Institution and the University of Arizona, on  data  from the NASA/ESA {\it Hubble Space Telescope\/}
obtained at the Space Telescope Science Institute, operated by the Association of Universities for Research in Astronomy, Inc., under NASA contract NAS5-26555, and on observations 
from the Gemini Observatory,  operated by the
Association of Universities for Research in Astronomy, Inc., under a cooperative agreement
with the NSF on behalf of the Gemini partnership: the National Science Foundation (United
States), the Science and Technology Facilities Council (United Kingdom), the
National Research Council (Canada), CONICYT (Chile), the Australian Research Council (Australia),
Minist\'{e}rio da Ci\^{e}ncia e Tecnologia (Brazil)
and Ministerio de Ciencia, Tecnolog\'{i}a e Innovaci\'{o}n Productiva (Argentina).} 

\altaffiltext{2}
{Minnesota Institute for Astrophysics, 116 Church St SE, University of Minnesota
, Minneapolis, MN 55455; roberta@umn.edu, kd@astro.umn.edu}

\altaffiltext{3}
{Department of Astronomy, The Ohio State University, 140 W. 18th Ave., Columbus, Ohio 43210}

\altaffiltext{4}
{Department of Astrophysical Sciences, Princeton University, Princeton, NJ 08544}

\altaffiltext{5}
{Department of Astronomy \& Astrophysics, Penn State University, 525 Davey Lab, University Park, PA 16802} 

\begin{abstract}
SN2011ht has been described both as a true supernova and as an impostor. 
In this paper, we conclude that it does not match some basic 
expectations for a core-collapse event.  We discuss SN2011ht's 
spectral evolution from a hot dense wind to a cool dense wind, 
followed by the post-plateau appearance of a faster low density wind 
during a rapid decline in luminosity.   We identify a slow dense wind 
expanding at only 500--600 km s$^{-1}$, present throughout the 
eruption.  A faster wind speed $V \sim 900$ km s$^{-1}$  
occurred in  a second phase of the outburst. There is no 
direct or significant evidence for any flow speed 
above 1000 km s$^{-1}$; the broad asymmetric wings of Balmer 
emission lines in the hot wind phase were due to Thomson scattering, 
not bulk motion.  We estimate a mass loss rate of order 
0.05 $M_\odot$ y$^{-1}$ during the hot dense wind phase of the event.  
The same calculations present difficulties for a hypothetical 
unseen SN blast wave.
There is no evidence that the kinetic energy greatly exceeded the 
luminous energy, roughly $3 \times 10^{49}$ ergs;  so the radiative 
plus kinetic energy was small compared to a typical SN.   
{\it We suggest that SN2011ht may have been a giant eruption driven 
by super-Eddington radiation pressure, perhaps beginning a few months 
before the discovery. A strongly non-spherical SN might also 
  account for the data, at the cost of 
  more free parameters. } 
\end{abstract} 


\section{Introduction}

Our multi-wavelength observations of SN2011ht have revealed 
an unusual eruption which shares characteristics with both the Type 
IIn supernovae and the SN impostors, Paper I (Roming et al 2012).  
The first spectrum of SB2011ht obtained shortly after discovery 
(Pastorello et al. 2011)  resembled the dense winds observed in some Luminous 
Blue  Variables (LBVs) at maximum, the Intermediate-Luminosity Red Transients 
(ILRTs) such as SN2008S and  NGC 300-OT2008 (Smith et al. 2009, Bond et al. 2009, Berger et al. 2009, Humphreys et al. 2011) and the warm hypergiant 
IRC+10420. With a  luminosity of Mv $\sim$ $-$14 mag at its discovery, SN2011ht  was 
designated an ``impostor'' (CBET 2851, PSN J10081059+5150570). Shortly afterward, 
however, it brightened  about two magnitudes in the visual, reaching  Mv $\sim$ $-$17 mag, at its distance of 19.2 Mpc  in UGC 5460 (Roming et al 2011). Based on its luminosity and narrow hydrogen  emission lines, Prieto et al. 
(2011a,b) suggested that it was a true supernova of Type IIn.  Simultaneous 
with its visual brightening, SN2011ht also showed  a rise in its UV flux,  increasing 7 magnitudes in 40 days (Roming et al. 2011). 
 SN2011ht is one of only a few Type IIn SNe observed in the near-UV and  
 has the most complete observations showing its rise to the UV and visual 
 maxima.

The first spectra obtained by our group after the rise in  UV flux were 
dominated by strong Balmer emission with P Cygni absorption and very broad 
asymmetric wings characteristic of Thomson scattering plus peculiar broad 
He I emission. It resembled a dense,  hot stellar wind 
(see Fig. 3 in Paper I). In Paper I we  attributed this peculiar spectrum 
to the interaction of the 
shock with the  ejection of a shell of material a year before the 
explosion. However, the last spectrum published in Paper I showed the onset of
a major spectroscopic change with the appearance of absorption and emission
lines characteristic of a much cooler wind as the luminosity  began to
decrease. A later spectrum from mid January revealed a 
fully developed dense wind resembling a late F-type or early G-type 
supergiant with strong absorption lines, plus H, the  Ca II triplet  and 
Fe II in emission with deep P Cygni profiles (Humphreys 2012). Shortly after that, 
SN2011ht began a rapid decline in the visual. A spectrum obtained shortly after the decline began shows another dramatic change; to a thin, warm wind characteristic of high mass losing luminous stars and some giant eruption LBVs. In many ways, SN2011ht closely resembles
SN1994W (Chugai et al 2004, Dessart et al. 2009), SN1998S (Chugai 2001), and 
SN2009kn a twin of SN1994W (Kankare et al. 2012). In this paper we emphasize that 
SN2011ht's development seems consistent with a radiation-driven eruption not a conventional 
hypersonic blast wave. The characteristic speeds are less than 1000 km s$^{-1}$.

The new observations including
near UV spectra from HST/STIS/MAMA and near and mid-infrared photmetry are described 
in the next section. The evolution of its spectrum and the kinematics of the ejecta 
are discussed in sections 3 and 4.
In section 5  we demonstrate that the asymmetric wings are consistent with Thomson scattering 
and use a scattering model to estimate the mass loss rate. The post-decline spectral energy 
distribution and dust formation properties are discussed in section 6.  The changing character 
and physical properties in 
the ejecta as the eruption progressed are reviewed in section 7. 
 In the last section we show that there is no observational evidence that SN2011ht was a terminal explosion and propose  an alternative  explanation for its  eruption.

\section{Multi-wavelength Observations and Data Reduction}

Our observations of SN2011ht range from the near ultraviolet beginning shortly after its
discovery to the near and mid-infrared at later times. 

\subsection{Ultraviolet, Optical and Near and Mid-Infrared Imaging and Photometry}
The {\it Swift} UVOT photometric measurements  are described in Paper I with a table of observations beginning only 5.8 days after discovery 
from JD 5838.6 to JD 5950.7, shortly before it began its rapid visual decline. The subsequent {\it Swift} measurements are in Table 1,  and in Figure 1 we show the 
complete multi-wavelength light curve from {\it Swift}. 
 Unfortunately no observations were
obtained by our group during the initial decline. 

The rapid decline by about 4.5  magnitudes in only 25 days was suggestive of 
dust formation. We therefore obtained JHK$_{s}$ photometry with the Lucifer 
near-infrared imager on the Large Binocular Telescope (LBT) on 13 March 2012 (JD5999.6). The seeing was $0 \farcs 9$, FWHM. A star in the field, 
2MASS 10081759+5151289, with magnitudes J = 10.83, H = 10.40, and Ks = 10.24, 
was used for the flux calibration. This star was slightly 
saturated in the images, complicating our data reduction. The flux was estimated by 
fitting the 2MASS star's image profile with a measured, scaled profile from a nearby
 fainter source for pixels greater than $0 \farcs 6$ (5 pixels) from the image center. 
 The flux was then computed using the area under this scaled profile. Our internal 
 errors were 0.02 mag or less at H and Ks, but due to uncertainties in our 
 computation of the comparison star's flux, we have included an additional error 
 of 0.1 mag in Table 2.

Visual ugri photometry was measured shortly afterwards on 25 March 2012 with
the MODS imager on the LBT with a 6$\arcmin$ $\times$ 6$\arcmin$ FOV and spatial scales of 
0.240$\farcs$/pixel and 0.246$\farcs$/pixel in the blue and red, 
respectively. The magnitudes were measured differentially with respect to seven 
isolated stars on the 
frame from the SDSS DR6 using aperture photometry with the QPHOT  routine in 
IRAF. The errors were determined by normal error propagation including the 
photometric errors  of the reference stars.  
The resulting spectral energy distribution (Figure 2) showed that SN2011ht was 
very red after the rapid  decline, with a rise in flux in the near-infrared, a probable  
signature of dust emission at longer wavelengths.  

We therefore requested and 
were granted Director's Discretionary Time on Gemini-N for mid-infrared 
observations at L$^{\prime}$ 
and M$^{\prime}$ with NIRI. Due to scheduling constraints and instrument availability the 
observations were not obtained until 03 May 2012. Images were acquired at 2.2, 3.8 
and 4.7$\mu$m. HD 84800 was used as a standard star with assumed magnitudes of K = 7.54, 
L$^{\prime}$ = 7.55, and M$^{\prime}$ = 7.56. Seeing was consistently under 
$0 \farcs 4$. A synthetic aperture of $0 \farcs 86$ was used for the standard
 and SN2011ht. For the L$^{\prime}$ and M$^{\prime}$ filters, images were taken in a 9 dither pattern, 
 spaced $3 \arcsec$ apart in RA and DEC. Sky background images were formed from
  the 8 other dither positions and subtracted from each dither position. 
  These 9 images were then shifted and stacked to produce the final image. 
  The $3 \sigma$ upper limit for the 4.7 $\mu$m photometry was determined from 
  the rms fluctuations in the final image and the size of the synthetic aperture.

The resulting visual and infrared magnitudes are summarized in Table 2 and the 
SED is shown in Figure 2. Although delayed, the L$^{\prime}$ and M$^{\prime}$ measurements confirm the formation of dust. The  spectral energy distribution and  
dust formation are discussed in \S {6}.

\subsection{Spectroscopy}

Spectra were observed with several different telescopes beginning shortly after
SN2011ht's discovery until it became too faint during the rapid decline.  
The observations described in Paper I include spectroscopy with the 
Hobby-Eberly Telecope (HET) LRS, the {\it Swift} UVOT uv grism, the 
Astrophysical Research Corporation (ARC) telescope DIS, and the Large Binocular
Telescope (LBT) MODS1. The details are decribed in Paper I. Additional groundbased  spectra  discussed in this 
paper were observed with the MMT Hectospec MOS on 23 January and 13 February 2012. The Hectospec 
spectra were obtained with the low resolution 270l grating yielding 
a wide wavelength coverage from 3650 -- 9200{\AA} with a FWHM resolution of 
5{\AA}. The
spectra were bias subtracted, flat-fielded and wavelength calibrated in the 
Hectospec pipeline at the Center for Astrophysics and flux calibrated. 

As discussed in Paper I, SN2011ht is distinguished by its rapid rise in the ultraviolet, strong UV flux, and a  steeply 
rising flux in its near-UV spectra. For these reasons we requested and were
granted Director's Discretionary Time with the {\it Hubble Space Telescope} 
for near UV spectroscopy with STIS/MAMA. Spectra were observed in three wavelength regions, NUV2800, NUV2376, and FUV1425 on 26 December 2011, and 
were processed through the calstis pipeline at STScI. Unfortunately these observations occurred just a few  weeks too late to see the UV-bright phase. Details of the 
observations are included in Table 3 together with the other spectra discussed 
in this paper. 

\section{Evolution of the  Spectrum}

The spectrum of SN2011ht has shown some remarkable and fairly rapid changes from its  initial ``dense wind'' at discovery, to the hot dense wind spectrum 
described in Paper I, and the transition to the cool dense wind observed later.
Our spectra showing the transion from the hot dense wind to the cool dense wind are shown
in Figure 3. 
Line identifications and the measured Doppler velocities with their 1 $\sigma$ errors
 are
summarized in Tables  4 and 5  at four significant 
times; the dense hot wind, the transition, the cool dense wind and  
a warm lower density wind   during the decline. 
 The spectra at these four 
times are discussed separately in the subsections below.   

\subsection{The Hot Dense Wind Spectrum}

The spectra obtained by our group immediately after the rise in the UV flux are shown 
in Paper 1. The spectra from that period correspond to the maximum in the UV flux. Our spectra observed between 01 and 23 November  2011  
show strong narrow Balmer emission lines with P Cygni absorption profiles
and 
 very broad emission wings. The wings are asymmetric to the red, a classic 
 signature of Thomson scattering (Figure 8). The  emission wings on the Balmer lines lines extend from
$\sim$  $-$4700 to $+$8000 km s$^{-1}$  in H$\alpha$, but, as discussed  in
\S {5} the broad wings are due to scattering
 by the electrons not bulk motion.
The two strong He I emission lines at 
 $\lambda$5876{\AA} and $\lambda$7065{\AA} are especially peculiar with very broad
asymmetric wings extending from -2600 to 5500 km s$^{-1}$ while the other He I lines 
show only weak emission
 and P Cygni absorption profiles similar to the hydrogen lines. Overall, the spectrum suggests a hot, dense wind, and in Paper I, Figure 9,
 a maximum temperature of $\sim$ 13,000 K was estimated for the wind.
 However the strong UV flux and the
 very steep rise in the near UV flux in the MODS1 spectrum are
not characteristic of a dense wind.The MODS1 spectrum from 
 17 November 2012 is shown here in Figure 4. Although the emission lines and 
 the 
 absorption features were listed in Paper I with their blueshifted 
 velocities relative to the galaxy,  due to a small zero
  point error in the wavelengths, they are repeated here
 in Table 4  with their measured Doppler velocities.

The P Cygni absorption components in the stronger Balmer lines are blueshifted
relative to the emission by about 550 km s$^{-1}$ supporting an expansion of the ejecta by approximately this amount.  We avoid using the FWHM of the Balmer emission lines to 
measure the velocity of the ejecta or wind because the line profiles are distorted on the red side by the strong scattering wings. The absorption components are likewise altered
by the strong emission. Therefore, as the emission weakens in the higher Balmer lines, the 
velocities of the Balmer absorption lines increase.  
 The spectrum also shows strong absorption 
from the O I triplet at $\lambda$7774{\AA}, a high luminosity indicator in A to F-type 
supergiants,  plus a weak  Ca II K line. 
 The O I and K-line velocities 
are comparable to those of the higher  Balmer absorption lines, not affected by
emission lines, indicating that these lines are  formed in the expanding ejecta. 
Although these lines are usually identified with somewhat lower temperatures, the O I triplet and a weak K line are observed in luminous supergiants with temperatures $\le$ 15,000K

Another peculiarity is the presence of three very broad absorption features in the near-UV in the  MODS1  spectra. These absorption features   
are unidentified, and are clearly present in both MODS1 spectra from November.  
They are not the atmospheric O$_{3}$  Huggins bands which are 
bluer and narrower and can be identified in the same spectra. They are not 
instrumental because they are not seen in the standard star spectra, and are
too sharp to be artifacts of the response. They  also do not correspond to any 
night-sky emission lines. Given their appearance, regular spacing, and sharp 
red edges, they are probably molecular. The wavelengths of the red edges are  $\lambda\lambda$3396,
3528, and 3659{\AA}\footnote{Correcting for the redshift of UGC 5460 of 1093 km s$^{-1}$, the corresponding wavelengths are 3384, 3515, and 3546{\AA}}.  We conclude that these absorption bands are real and are most likely formed 
near SN2011ht and probably in the ejecta. They could be diffuse interstellar bands (DIBs), but none are known at 
these wavelengths.

\subsection{The Transition}

The last spectrum described in Paper I   from 21 Dec. 2011 (see Figure 3) shows the onset 
of the transition from the hot wind to the cool dense wind as the total luminosity and the 
near-UV flux  declined during this time.  The MODS1 spectra from  
26 and 28 December 2011 (Figure 5) show a well developed spectrum of Fe II absorption and emission 
lines, Ca II H and K absorption, and the appearance of the near-infrared Ca II triplet
in absorption.  The Balmer lines still show  
the strong asymmetric emission profiles with P Cygni absorption. The He I emission is gone, but an
excess of flux is still present in the near-UV plus a strong forest of Fe II absorption lines.

 We have four spectra from this period obtained only a few days apart that illustrate  the relatively rapid transition. There are no apparent P Cygni absorption components in the Balmer lines in the HET 21 December spectrum, although they  are clearly visible 
 in the MODS1 spectrum a few days later.  This difference is due to the lower resolution in the HET spectrum. In this series of spectra,  the Fe II absorption lines increase in strength with time, 
 the O I $\lambda$7774 and $\lambda$8446 absorption lines increase,  and the Ca II triplet 
 appears. Between 21 December to 02 January the equivalent widths of the Ca II lines 
 nearly doubled. 
 Similarly, the O I triplet and the  Fe II absorption lines in the blue increased by about 
 a factor of 1.5 in equivalent width.

\subsection{The STIS/MAMA Ultraviolet Spectra}

As we noted earlier, SN2011ht's UV flux and steep continuum in the near-ultraviolet in the MODS1 spectrum from 17 November 2011 are not characteristic of a
dense wind, even a very hot one. The strong UV flux might be due to shock breakout from the dense circumstellar medium as suggested for PTF09UJ (Ofek et al. 
2010), although we would expect the UV peak to be brief and the temperature 
to drop not rise after the breakout. Consequently, these features raise questions about how the object  can
produce such a steep near-UV continuum if the wind is opaque and if it 
isn't opaque, 
how does it produce a continum with a supernova-like brightness plus emission 
lines with such strong Thomson scattering (see \S {5} below). To answer these questions we obtained
near-UV spectra shortward of 3300{\AA} with STIS/MAMA for 
independent information on its temperature and other parameters.  Unfortunately,
the spectra were obtained just after the UV flux had begun to decline and
when the groundbased spectra showed the appearance of the absorption and emission lines of
Fe II signalling a decrease in temperature of the wind. 

 The MAMA spectra were observed 
on the same date as the MODS1 spectrum from 26 December 2011. The NUV2376 
spectrum and MODS1 blue spectrum almost overlap and are shown together in 
Figure 5.  Both spectra show a forest of absorption lines (Fe II, Cr II, Mn II) shortwards of the Balmer jump common in the spectra of  supergiants with temperatures 
below about 12,000K. Other strong features in the near UV spectra include the Mg II doublet at $\lambda$2800{\AA} in absorption. The absorption lines of Si II, Si III, and N III 
reported in the earlier {\it Swift} uv spectra (Paper I) are not present.

\subsection{The Cool Dense Wind}

Figure 6  shows the Hectospec spectrum for 23 January 2012 together with the MODS1 26 December 
spectra for comparison. The 23 January spectrum shows the fully developed cool, dense wind. Although SN2011ht was still visually bright it was fading in $b$ and $u$. 
The absorption line spectrum resembles that of  an early  G-type star with a strong G band plus high 
luminosity features such as the blends near $\lambda$4172{\AA} and $\lambda$4177{\AA}.
The Balmer lines are still in emission with broad asymmetric wings but with 
noticeably weaker emission,  and the 
P Cygni absorptions are much deeper relative to the continuum. The Ca II triplet absorption lines  have 
continued to increase in strength and now show strong P Cygni profiles with broad emission wings 
asymmetric to the red like the Balmer lines. The \ion{Ca}{2} triplet emission lines are 
formed in the star's ejecta   by radiative de-excitation from the strong  \ion{Ca}{2} H
and K absorption upper levels.  \ion{Ca}{2} has a low ionization potential and is 
suppressed in the presence of UV radiation consequently, the emission only began to weakly appear in 
the transition spectra with the decline of the UV flux. 
The Fe II absorption lines are 
also stronger and their corresponding emission features when present are much weaker. We also note a strong Na I D line with a P Cygni profile.   

\subsection{The Decline -- a Low Density Wind} 

We have three spectra observed in February 2012 during SN2011ht's rapid decline in 
apparent brightness. The spectra from 13 February and 28 February, 2012 are shown in Figure 7.  Although, all three exposures are weak with low S/N, they  are essentially identical.  The strong absorption line spectrum is gone. The spectra show  
H$\alpha$, H$\beta$, H$\gamma$, the Ca II triplet, and weak 
[Ca II] $\lambda\lambda$7291 and 7323 in emission. Four  Fe II emission lines are 
also identified. The Na I D doublet 
still has a P Cygni-type profile. Although the Ca II triplet no longer 
shows absorption components, the emission lines are 
still noticeably asymmetric to the red. The scattering wings on the Balmer 
lines however are gone, but
 H$\alpha$ and H$\beta$ still have  weak P Cygni-type  absorption. The only absorption lines, not altered by emission, are H$\delta$ and H$\epsilon$ and they have velocities consistent with the absorption lines measured at the earlier times.
The presence of the rare  [Ca II] emission doublet indicates  a very low density optically 
thin region.  The transition  that produces the triplet emission
 leaves the ions in the upper level for the [\ion{Ca}{2}] forbidden lines. These
 ions are  normally collisionally de-excited back to the ground state that
 produces the H and  K lines, unless the density is sufficiently low. 

Our spectrum from 13 February was
observed   when the object was about 18th magnitude (Figure 1) and had decreased by 
2.5 magnitudes.  The wind had thus transitioned from its cool optically thick
state to a  much less dense state in only 20 days.
 The spectrum at this time resembles the winds of some luminous, 
  hot supergiants with high mass loss, and the spectra of some giant eruption LBVs or impostors 
 such as SN2009ip (Smith et al 2010, Foley et al 2010). 

The mean emission and absorption velocities for the four  phases  discussed here are summarized in Table 6.

\section{Kinematics of the Ejecta}

The measured Doppler velocities from the different emission line species (H, He I,
Fe II), at any given phase,   do not show any significant relative
offsets.  They are also 
consistent, within the errors, with the redshift of UGC 5460 at 1093 km s$^{-1}$ throughout the eruption, although the Balmer emission lines show a progression to higher velocities
with time, discussed below. There is no rotation curve for UGC 5460, so we do not know the expected velocity for SN2011ht. For this discussion, we therefore adopt the
galaxy redshift  as the intrinsic velocity for SN 2011ht.
The velocities of the absorption lines, excluding the
P Cygni-type  absorption components altered by emission,  are very uniform
throughout  the eruption with  mean velocities  at each phase  of $\sim$  530 -- 560  km s$^{-1}$.  The expansion  velocity of the circumstellar ejecta is thus $\sim$ 550 km s$^{-1}$, blueshifted  relative to SN2011ht. Given
the uncertainty in the intrinsic velocity, the   expansion  velocity is 
realistically between 500 and 600  km s$^{-1}$.  

Only during the decline, when the ejecta had become optically thin,  do we see a
significant change in the velocities. 
Both the Ca II triplet and the [Ca II] emission lines
show a statistically significant difference of $\approx$ 150 km s$^{-1}$
and 350 km s$^{-1}$, respectively,  with respect to the
hydrogen emission. As already mentioned the  [Ca II] lines are formed in a very low
density region which is thus moving outward relative to the object or the hydrogen
emission region.

The velocity difference between the emission peaks and ther corresponding  
absorption minima
appears to increase with time from the hot wind stage to the decline, 
from about 560 km s$^{-1}$ in the hot wind and transition phases to  640 km s$^{-1}$
in the cool wind stage in both the hydrogen and the Ca II lines. The  weak
absorption mimina in  H$\alpha$ and H$\beta$ in the  low density wind have a velocity
difference as  high as 700 km s$^{-1}$ relative to the emission peaks. However,
inspection of the velocities in Table 5 shows that this is actually due to an
increase in the redshift of the emission lines suggesting a possible contraction
of the ionization zone.

In the next section we   
show  that the broad asymmetric wings  are consistent with Thomson scattering and use the 
scattering model to estimate the outflow density and mass loss rate.

\section{The Scattering Wings}

The Balmer emission lines have the familiar characteristics of Thomson
scattering in a dense outflow: roughly triangular profiles with concave sides,
asymmetric to the red, and with wings extending much farther than one would
expect from the absorption line velocities. The individual profiles for the H$\alpha$, H$\beta$
and H$\gamma$ are shown in Figure 8.  The Doppler-shifted 
wings  do not indicate bulk velocities, but are instead caused by
multiple scattering events.  Thomson scattering  has at least two major implications:
(1) There is no persuasive evidence for any outflow velocities much
greater than about 600 km s$^{-1}$ during the eruption, and (2) the line wings can
be used to estimate the mass loss rate (see below).
Additional support for this interpretation is presented in \S {5.1}.

Dessart et al. (2009) convincingly demonstrated that  similar asymmetric
wings in SN1994W were due to Thomson scattering. 
Figure 11 in their paper, for example, shows a Thomson scattering model profile that
closely resembles our Figure 8.
For SN2011ht the strength of the scattering wings compared to 
     line-center increases mildly from H$\alpha$ to H$\beta$ to H$\gamma$;  
     qualitatively this effect is expected but it seems  less conspicuous 
     than in the case described by Dessart et al.\ (cf.\ their Fig.\ 12).
The wavelength offsets and velocities corresponding to the Thomson scattering
wings are summarized in Table 7 for H$\alpha$ and other lines.
In SN2011ht, the wings also appear to become  less extended  at later times,
presumably due to a decrease in the electron density and/or temperature.
Since the ejecta are optically thick, we attribute this change primarily
to a decrease in the apparent temperature and in $\tau_{sc}$ as the eruption
progressed from the hot to cool wind phase. There are no apparent
Thomson scattering wings on the H$\alpha$ and H$\beta$ emission profiles
observed in the warm, thin wind during the decline when $\tau_{sc}$ is 
expected to be small.

\subsection{The Balmer lines and the flow density} 
\label{ssec:balmerwings}  

Adopting the Thomson scattering model for the broad wings in the Balmer 
lines, we can obtain simple order-of-magnitude estimates of the outflow 
densities and the mass loss rate.  We employ a few very simplified  
assumptions that  most likely resemble the true situation.   When 
similar reasoning was applied  to the case of $\eta$ Car's wind 
(Davidson et al 1995), the results were  close to those found later with 
an  elaborate wind model (Hillier et al 2001).   

Consider a steady, constant-velocity spherical mass outflow with locally 
averaged electron density ${\langle} n_e(r) {\rangle} \, = \, \zeta/r^2$, 
where $\zeta$ is a constant parameter which we hope to estimate.    
In the simplest reasonable model, the emission line in question 
originates outside some radius $r_1$, while at smaller radii the photons are 
presumably absorbed by continuum processes before they escape.  In this 
case the emergent emission-line luminosity can be represented as 
a volume integral 
  \begin{equation}  
    L_\mathrm{line} \ \approx \  {\int} \,  A \, n_e^2 \, dV  
        \ \approx \  \frac{4 \pi  A \, \zeta^2}{\epsilon \, r_1 } \; ,  
  \label{eq:llum}  
  \end{equation}
where $\epsilon \leq 1$ is the standard ``clumping factor'' to allow for 
inhomogeneous densities. The emission rate parameter $A$ is roughly 
known for each Balmer recombination line (Osterbrock \&  Ferland, 2006), 
except that continuum absorption  and other effects  reduce its effective 
value in a dense wind.    
Meanwhile the Thomson scattering optical depth at radius $r_1$ is 
  \begin{equation}
    {\tau}_\mathrm{sc} \ \approx \ \frac{{\sigma}_e \, \zeta}{r_1} \, .
  \label{eq:tausc}
  \end{equation}
If observations give estimates of $L_\mathrm{line}$ and ${\tau}_\mathrm{sc}$, 
we can eliminate $r_1$ to find the density parameter: 
  \begin{equation} 
    \zeta \ \approx \ 
         \frac{{\sigma}_e \, \epsilon \, L_\mathrm{line}}
              {4 {\pi} A \, {\tau}_\mathrm{sc}} \, .
  \label{eq:qeqn}
  \end{equation} 
With an adopted outflow speed, this gives an estimate of the mass 
loss rate.  Various subleties which do not 
alter the basic reasoning have been omitted here. 
Note that the average emergent line photon originates around 
   $\tau \, \sim \, 0.5 \tau_\mathrm{sc}$.

In this calculation we adopt the parameters measured for H$\beta$ in the  MODS1 spectrum from 
2011 November 17.  A Thomson-scattered line profile is broadened and 
shifted by the electrons' thermal motions (e.g., $v_e(\mathrm{rms}$) 
$\approx$ 830 km s$^{-1}$ at $T = 15000$ K) and by expansion of the 
outflowing gas.  We can estimate ${\tau}_\mathrm{sc}$ from the average 
number of scattering events before escape, which is indicated by the 
line broadening.  Unfortunately the  line width is not easy to parametrize 
in terms of observables, because the wings have no definite limits and 
the ill-defined, unscattered core of the line affects its central part.  
An acceptable measure for our purposes is 
  \begin{equation}
     U_\mathrm{sc}  \ = \   
     \frac { \int | {\lambda} - {\lambda}_c | \, F({\lambda}) \, d{\lambda} }
           { \int F({\lambda}) \, d{\lambda} } \; , 
  \label{eq:defwidth} 
  \end{equation}  
where ${\lambda_c}$ is the centroid of the emission line.\footnote{  
   %% FOOTNOTE
   The r.m.s.\ width, using $({\lambda} - {\lambda}_c)^2$ instead of 
   $| {\lambda} - {\lambda}_c |$, is theoretically more apt but in practice is 
too sensitive to the poorly  
   measured extremeties of the wings. }  
Thus we estimate $U_\mathrm{sc} \approx 1850$ km s$^{-1}$ for H$\beta$ 
in 2011 November.  
  If $v_{e}(\mathrm{rms}) \approx 830$ km s$^{-1}$ and the wind speed 
  is 550 km s$^{-1}$ based on the absorption features, then the 
  estimated value of $U_\mathrm{sc}$ requires 
  $\tau_\mathrm{sc} \, \sim \, 3.5$.  This estimate is based on Monte 
  Carlo trials with the source distribution stated in eq.\ 1.  
  It is inexact because the assumptions are simplified, and the central 
  part of the line shape is not modeled realistically.  The line wings, 
  however, are represented better and would appear visibly different 
  if $\tau_\mathrm{sc}$ were less than 1.8 or more than 7.  In the 
  former case, the long-wavelength wing would be only a relatively weak 
  satellite to the core of the line; while in a high-$\tau_\mathrm{sc}$ 
  case the profile would be  dominated by the 
  wings with no perceptible core.  The range 
  $1.8 \lesssim \tau_\mathrm{sc} \lesssim 7$ spans a factor of two 
  in $U_\mathrm{sc}$.   

  The luminosity of H$\beta$ in November 2011 was about $1.6 \times 10^{40}$ 
  ergs s$^{-1}$.  The parameter $A$ in eq.\ 1 would have a value of  
  about $10^{-25}$ erg cm$^{3}$ s$^{-1}$ in a nebular case
  \citep{OF}, but absorption in the line formation 
  region can somewhat reduce the effective emissivity;  we tentatively 
  adopt the value just quoted.  Putting these quantities into eq.\ 3, 
  we find $\zeta \, \sim \, 2 \times 10^{39} \, \epsilon$ cm$^{-1}$, 
  which implies a mass-loss rate of the order of 
  $0.05 \, \epsilon \ M_\odot$ y$^{-1}$ if 
  $V_\mathrm{wind} \approx  550$ km s$^{-1}$.  These values imply  
  $r_{1} \, \sim \, 30\epsilon$ AU, compared to a photosphere size
  of roughly 30 AU based on the continuum brightness (Paper I and 
  {\S 7} of this paper).  In view of the uncertainties, this 
  order-of-magnitude outcome appears satisfactory.   In order to 
  prevent $r_1$ from becoming implausibly small, evidently the  
  clumping factor $\epsilon$ cannot be very small.  Note that 
  {\it our $r_1$ estimate for November 2011 is almost entirely 
  independent of the photospheric size estimate, and it turns 
  out to have the same order of magnitude.}

Another independent argument also appears consistent with the deduced 
mass loss rate.  Davidson (1987) assessed the relation 
between mass loss rate and photosphere radius in an opaque wind with 
a given luminosity.  Figure 1 in that paper is a plot of photospheric 
temperature vs.\ a relevant quantity $Q$ which depends on mass loss 
rate, wind speed, and luminosity.  The parameters quoted above for 
SN2011ht imply $\log Q \, \approx \, -5.5$, but in order to compensate 
for modernized opacities we should increase this to roughly $-5.2$ in that figure.  
{\it This value places 
the outflow near a critical part of the temperature vs.\ $Q$ 
relation.\/}  It is consistent with  a photospheric temperature fairly close 
to $10^4$ K as observed; and, more important, it also implies that 
either a moderate decrease in luminosity or a moderate increase in the 
mass loss rate would change the spectrum to a classic ``cool dense wind'' state. 
 The character of the  spectrum did indeed change in that way 
a few weeks later (\S {3.4}).   In summary, two very different 
modes of reasoning are both consistent with a mass loss rate of 
the order of 0.05 $M_\odot$ y$^{-1}$ in 2011 November;     
  and they are both consistent with the size of 
the continuum photosphere inferred from its brightness and 
color.

The Thomson broadening argument gives a less satisfactory result 
if we concentrate on H$\alpha$ rather than H$\beta$.  With a 
nebular value of the emissivity $A$, H$\alpha$ seems to imply 
$r_1 \, \lesssim  \, 20$ AU.  However, self-absorption is  
much stronger for H$\alpha$ than for the other Balmer lines;  
so the effective value of $A$ in the emission zone is likely 
to be substantially smaller than the nebular value, thereby 
eliminating the discrepancy.    
In any case, this discrepancy is no worse than 
the factor-of-two uncertainty expected for our simplified 
approximations. Higher Balmer lines appear satisfactory.

The size scale noted above, of the order of 30 AU for both 
continuum and emission lines, presents difficulties for any 
model with a massive supernova shock  wave.  The reasons will be outlined 
in {\S}\ref{sec:discuss} below.

Radiation pressure exceeded the Eddington Limit by perhaps two orders 
  of magnitude in November 2011,  but this has no bearing on the 
  emergent-radiation analysis. Altogether, the case for 
  Thomson scattering line wings is very strong.   The observed 
  profiles have the right shapes and widths (cf.\  Fig.\ 11 in Dessart 
  et al.\ 2009),  and the analysis based on this view leads to 
  the same wind density as an entirely independent mode of 
  reasoning as stated above.  Moreover, the characteristic size of the 
  relevant zone is found to be of the same order of magnitude as 
  the photosphere estimated in Paper I.  Any alternative hypothesis 
  would need to explain all these facts.   If SN2011ht  was a true 
  supernova, then it must have generated the protracted post-explosion 
  low-speed dense outflow described above; and any faster shock wave  
  must have escaped observation.  If, on the other hand, SN2011ht was 
  a different type of eruption, then the low-speed flow may have been 
  the main event.  Dynamical models of such an extreme super-Eddington 
  case are beyond the scope of this paper.

Although the above reasoning is highly simplified,  a full-fledged 
wind model like Dessart et al.\ (2009) may not improve the 
results as much as one might hope.  The velocity and 
inhomogeneity functions $v(r)$ and ${\epsilon}(r)$ are scarcely 
known for a continuum-driven eruptive outflow; radiative transfer 
in the Balmer lines may have intricate local variations; the 
example of $\eta$ Car shows that spherical symmetry can be a poor 
approximation, etc.  In principle, these effects together imply 
that a standard wind-code model is valid for this type of object 
only in an order-of-magnitude sense -- i.e., its derived quantities 
are only moderately more accurate than our simple approach.

A final point: The spectra of several Type IIn supernovae show similar broad asymmetric wings in the 
Balmer lines (Kiewe et al 2012) which have been interpreted as due to actual mass motion,
a mix of expansion and scattering, and as pure Thomson scattering. The wings in SN1994W 
have been demonstrated to be due to Thomson scattering, and in SN2011ht we have also 
shown that the behavior of the asymmetric wings is consistent with the expectations of Thomson 
scattering. {\it Therefore a word of caution is justified  regarding 
 similar asymmetric broad wings observed in several Type IIn supernovae. Fitting 
these profiles  with  multiple Gaussians or other arbitrary distributions is very 
likely to  lead to erroneous conclusions regarding the expansion velocities.} 

\section{The Later Spectral Energy Distribution and Dust Formation } 

SN2011ht's visual brightness declined by 4.5 magnitudes in less than 
25 days during January and February 2012 (Figure 1), and afterward  significant  flux was measured at 
$\lambda \sim$ 2 to 4 $\mu$m (Figure 2).  This near and mid-IR emission 
most likely indicates the presence of warm dust.
In addition to the rapid decline, the visual photometry obtained at the end of the plateau phase (JD 5950.7, Table 1)
compared with the post-decline photometry in Table 2 suggests that the object also got 
significantly redder. Its $b-v$ color of 0.65 mag immediately prior to the decline 
coincides with the cool wind phase and is consistent with the apparent spectral type and the 
low reddening derived in Paper I.  The $ugri$ photometry measured 60 days later is equivalent to a redder $b-v$ color of $\sim$ 0.95 mag\footnote{The $r$-band magnitude is contaminated by  H$\alpha$ emission (see Figure 2). Correcting for the H$\alpha$ flux gives  a slightly
fainter $R$ magnitude of 20.1. Converting the ugri photmetry \citep{Kar05,Jest05} yields 
 a corresponding $b-v$ color of 0.95 - 0.99 mag.}  and a color excess of $\sim$ 0.3 mag, relative to the pre-decline colors. This could be due either to dust formation or a continued shift 
in the 
energy distribution to cooler temperatures. Unfortunately, the only spectra we have for 
comparison during the decline show an unresolved continuum and emission lines (Figure 7). 
Assuming that the color excess is all due to dust formation with a 
normal reddening law, A$_{v}$ is $\sim$ 1.0 mag, with R $=$ 3.2 and  could 
account for about one-fourth of the decline in brightness. 
The color excess also implies a neutral hydrogen column density, n$_{HI}$, of $\approx$ 2.6 $\times$ 10$^{21}$. Unfortunately one cannot make a trustworthy estimate of the mass of 
     the dust, because that would require several assumptions that are  
     very doubtful in this case.  (A precedent might be the $\eta$ Car 
     eruption, for instance;  its grains and value of $R$ are much larger 
     than normal and the reasons are not understood, see references in 
     \citealt{DH97}.)

Dust has been confirmed in several Type IIn's 
which somewhat distinguishes them from other supernovae 
(Fox et al. 2011, Otsuka et al. 2012).  Similar objects that declined 
rapidly and then were  observed later to have formed dust include SN2007od, SN2009kn, 
and SN1994W (Sollerman et al. 1998, Andrews et al. 2010, Kankare et al. 2012), all similar to SN2011ht.  
In these cases the rapid luminosity decrease  
was said to imply a very  low mass of radioactive $^{56}$Ni in the ejecta. 
Few authors take this idea to its logical limit:  perhaps   
the event was not a core-collapse SN, hence no $^{56}$Ni (cf. \citealt{Dess08}).  
Thus the presence of dust is potentially significant and the  near-IR flux is worth examining, 
even though 
theoretical difficulties preclude any satisfying model for the 
dust with the limited information we have.  

Only a small fraction of the observed near-IR flux can be thermal 
free-free emission.  In order to match the $\lambda \; \sim$ 2 $\mu$m 
flux (Figure 2) with this mechanism, a mass loss rate of the 
order of 0.01 $M_\odot$ y$^{-1}$ is required. Intuitively this rate 
seems higher than one would expect for 
      January 2012, since it is nearly of the same order of magnitude as the 
      value estimated in {\S}5 above for an early, much more active 
      stage of the event.  Concrete objections to the free-free 
      hypothesis are as follows. The resulting 
${\lambda}F_{\lambda}$ would rise toward smaller wavelengths more 
than was observed, and the emission lines should be much stronger 
than those shown in Figure 7.  Thus we can reasonably assume 
that the 2--4 $\mu$m flux originates chiefly from scattering and emission by  
warm dust grains 
rather than thermal bremsstrahlung.  The central object's continuum
plus bremsstrahlung may account for an appreciable fraction at
$\lambda \; \lesssim$ 2 $\mu$m, however. 

If, for example, 70\% of the 2 $\mu$m flux and all of the 4 $\mu$m 
flux were emitted by grains, then the 2--4 $\mu$m spectral slope 
in Figure 2 corresponds to a dust temperature of roughly 900 K -- 
presumably an average over a range of grain temperatures.\footnote{   
   %%% FOOTNOTE
   With a typical absorption efficiency function, the wavelength 
   distribution of emission from a small grain at $T = 900$ K 
   resembles a 1100 K blackbody.  We mention this because some 
   authors have used Planck curves to represent observed 
   dust SED's. } 
Figure 2 indicates an integrated luminosity of about 
$6 \times 10^6 \; L_\odot$ for SN2011ht in March 2012, about 
$100\times$ less than it was three months earlier.  With this  
value, a grain absorption efficiency proportional to 
${\lambda}^{-1}$ (Draine 2011), and a source color 
temperature around 6000 K,  the distance from the star corresponding 
to dust temperature $T_d$ is 

%%   \begin{equation}  
%%      r  \  \approx  \  (500 \; \mathrm{AU}) 
%%         {\left(} \frac{1000 \; \mathrm{K}}{T_d} {\right)}^{2.5}  \; .  
%%   \label{eqn:dusttemp}  
%%   \end{equation}  

  \begin{equation}  
    r  \  \approx  \  
    ( 500 \; \mathrm{AU} ) \; ( T_d / 1000 \; \mathrm{K} )^{-2.5}  \, .  
  \label{eqn:dusttemp}   
  \end{equation}  

If the stated assumptions are close to reality, the observed near-IR
in Figure 2 originated mainly at locations 400 to 1000 AU from the 
central object.   At a speed of 550 km s$^{-1}$, the material in 
question would  have been ejected 3--10 years before the event. 
This idea seems basically reasonable, since a pre-event mass loss 
rate less than $10^{-3} \; M_\odot$ y$^{-1}$ would have given a 
sufficient column density in the region $r \, \gtrsim \, 400$ AU.

However, the same reasoning leads to a grain-formation puzzle.   
During the high-luminosity stages of the event, dust should have  
been heated to temperatures roughly $2.5\times$ higher, averaging 
about 2300 K instead of 900 K -- probably hot enough to destroy 
pre-existing grains.  Thus, at first sight, it seems likely 
that new grains formed as the luminosity declined in early 2012.
But at a gas density of the order of $10^{-19}$ or $10^{-18}$ g cm$^{-3}$ 
(consistent with the above remarks), a grain cannot grow to 
a size of, say, 0.01 $\mu$m in only 2 months via simple accretion 
of thermal atoms \citep{csk}. The shortfall in growth rate is two or three orders 
of magnitude.  We do not have a satisfying answer to this paradox, 
but note the following considerations:  
  \begin{enumerate} 
  \item 
  One can speculate that the relevant gas at $r \sim 500$ AU was compressed 
  by a shock wave plus subsequent cooling e.g., in a true SN explosion.  However, in order to 
  reach the radius of interest only half a year after the initial 
  explosion, the shock speed must have been several thousand km s$^{-1}$, 
  which implies a post-shock temperature above $10^8$ K and a cooling 
  time of many years in the relevant gas -- so the compression factor 
  would not have been very much larger than the adiabatic value 4, insufficient to explain 
  the phenomenon, while the high temperature would create an environment 
  hostile to dust.  In any case, deducing a strong shock wave from the 
  near-IR flux alone would be an extreme extrapolation.  
  \item
  It is not certain that all the relevant pre-existing grains were 
  destroyed by the maximum luminosity.  The 2012 near-IR flux might 
  have been produced by slightly cooler dust than we estimated above, 
  and parts of especially refractory grains may have survived up to 
  2000 K or so. The unidentified bands (\S {3.1}) may be relevant to this question. They 
  could be remnants  of previous existing circumstellar material.  
  \item 
  The grains might grow faster than the very simple estimate quoted 
  above.  If the pre-eruption wind speed was of the order of 100 km s$^{-1}$ 
  rather than 550 km s$^{-1}$ (i.e., like a cooler hypergiant or an 
  LBV ar maximum), or if the outflow is inhomogeneous, then local 
  density maxima at $r \sim 500$ AU could have greatly exceeded 
  $10^{-18}$ g cm$^{-3}$.  Meanwhile, radiation-driven motion of 
  the grains through the gas might cause faster growth;  and of
  course there may be other processes that we have not thought of.
  \item 
  Consider the near-IR flux of $\eta$ Car, a well-observed object 
  with practically the same luminosity and wind speed that SN2011ht 
  had in March 2012.   In 1970--2000 $\eta$ Car's  mass loss rate was of the 
  order of $10^{-3} \; M_{\odot}$ y$^{-1}$ (Davidson \& Humphreys 1997, 
  Hillier et al.\ 2001).\footnote{  
     %%% FOOTNOTE
     We cite old estimates because $\eta$ Car's wind density 
     has considerably decreased since 2000 
     (Mehner et al.\ 2010, 2012; Martin et al.\ 2006). } 
  Based on simple accretion of thermal atoms, one would predict that 
  the radius of a grain 500 AU from $\eta$ Car would initially grow 
  no faster than about $3 \times 10^{-4}$ $\mu$m y$^{-1}$, and 
  this rate would decrease within 3 years as the grain moves outward.  
  This is too slow to produce much near-IR emission in 
  $r \, \lesssim \, 1000$ AU.  In fact a 2--5 $\mu$m continuum was 
  observed and we can interpret its spectral 
  slope, as follows.  First consider an idealized spherical, 
  constant-velocity outflow.  Suppose that the absorption coefficient 
  of dust in this outflow has radial dependence 
  $k(\nu,r) \, \propto \, r^{-\alpha}$.  (Thus $\alpha = 2$ 
  means that the extinction per unit mass is constant.)     
  If $T(r) \, \propto \, r^{-0.4}$ and $kr << 1$, one can show that    
  near-IR emission from the dust has frequency dependence   
  $f_{\nu} \, \propto \, {\nu}^{2.5\alpha - 3.5}$.  The observed 
  2--5 $\mu$m spectrum of $\eta$ Car in 1970--2000 closely resembled 
  $f_{\nu} \, \propto \, {\nu}^{-3.8}$ (see, e.g., Figure 3 in  
  Cox et al.\ 1995).  Roughly speaking, this suggests 
  $k \propto r^{0.1}$ at radii $r \, \approx$ 500 to 2000 AU. 
  In other words, the amount of dust extinction per unit overall 
  mass appears to have grown substantially in a flow time of 10 years 
  or even 5 years -- contrary to the simple expectation  
  quoted above.  Evidently the well known example of $\eta$ Car 
  presents the the same near-IR puzzle as SN2011ht.  
  One can imagine  potential pitfalls 
  in this analysis, but it illustrates that the theoretical 
  grain-formation problem is not strong enough to invalidate  
  our basic view of SN2011ht's post-eruptive state.    
  The same problem exists even if the event was a true supernova. 
  \end{enumerate}   %%%

\section{The Stellar Wind, Circumstellar Ejecta, and  Mass Loss}

SN2011ht is remarkably similar to SN1994W and to SN2009kn which has recently been 
described as a spectroscopic twin to SN1994W (Kankare et al. 2012). These objects and 
similar Type IIn's have been assumed to 
be true supernovae of either Type IIn or Type IIP in which the shock from the terminal 
explosion interacts with a previously existing dense circumstellar medium. 
Dessart et al. however have suggested that SN1994W's spectroscopic evolution and luminosity can 
just as well be explained as the collision of two shells or ejections.  
The principle difference 
in the observational record between SN2011ht and  these  other
Type IIn's is the {\it Swift} near-UV photometry for SN2011ht. If  UV observations had 
been available for SN1994W, we suspect that  it likewise would have had a strong UV flux in its 
early hot wind phase. 

In this section we begin by discussing  the physical parameters of SN200ht's wind or 
ejecta that can be derived from its spectra and light curve.   

The discovery spectrum described by Pastorello et al (2011) showed a forest of Fe II absorption lines, Ba II and Ca II H and K in absorption, and hydrogen and the Ca II triplet 
in emission with P Cygni profiles. Although we do not have a copy of the spectrum, this description is consistent with the ``F-type" supergiant spectrum observed in 
several giant eruption LBVs and in the warm hypergiants. Adopting an apparent temperature of $\sim$ 7000 K with SN2011ht's luminosity of M$_{v}$ $=$ $-$14.4 mag at that time, the ejecta or dense wind had a radius of only 20 AU. If this 
material is expanding at 550 km s$^{-1}$, it would have reached this distance 
$\approx$ 0.2 yr prior to the discovery.

During the hot wind phase when the UV flux was a maximum in mid-November 2011, 
the apparent temperature of the dense wind was $\sim$ 13000$\arcdeg$K (Paper I).
The total luminosity at that time was 10$^{9}$ L$_{\odot}$ and the corresponding
 photospheric radius was 30 AU.
In \S {5.1}, we  estimated a  mass loss rate on 
the order of 0.05 $M_\odot$ y$^{-1}$ from a model for the scattering wings in the 
Balmer lines in this phase. 
This is much below what we would expect for a terminal
explosion and much higher than in the classical LBVs 
at maximum  which typically have  mass loss rates of 10$^{-4}$ to 10$^{-5}$ $M_\odot$ y$^{-1}$ (Humphreys \& Davidson 1994).

As the total luminosity declined and the UV flux decreased, the apparent
temperature of the wind decreased fairly rapidly 
as observed in the transition spectra, and the  photosphere moved outward into 
the ejecta.  In the cool dense wind phase the absorption lines correspond
to a late-F/early G-type supergiant. Adopting an apparent temperature of $\sim$ 6000 -- 6500$\arcdeg$K  
and a luminosity of 1.6 $\times$ 10$^{8}$ L$_{\odot}$, the photospheric radius was 
50 - 60 AU. With an expansion velocity of  550 km s$^{-1}$, the ejecta would have reached 
this distance in about 6 months. The temperature was very likely higher in the preceding warm phase because the dense ejecta had not moved far enough out for a cool photosphere.

The expansion velocity  measured from the
absorption lines throughout the eruption is $\sim$ 550 km s$^{-1}$ with some material up to 
$\sim$ 700 km s$^{-1}$.
The corresponding times for the ejecta to reach the above radii imply
pre-existing ejecta and an earlier or on-going eruption only shortly
before the discovery. The pre-outburst upper magnitude limit from Boles (2011) 
at $\sim$ 19.5 mag in April, 2011, also restricts  the previous outburst to 
no more than than 0.5 yr before the discovery, unless the initial outburst had already begun
at that time.

The transition to the thin wind stage during the decline
was accompanied by the appearance of the [Ca II] lines which are relatively rare
and require very low densities. Furthermore, the Balmer emission lines  
show no evidence of the Thomson scattering wings which had dominated their 
line profiles in the earlier spectra. Both of these observations suggest
that the optical depth and electron density greatly decreased. 
Since the hydrogen emission line profiles are no longer distorted by the strong wings, 
the  FWHM of the H$\alpha$ emission line measured at 900 km s$^{-1}$ may be indicative 
 of the 
corresponding wind speed. Although the few absorption lines present in the spectrum still indicate an outflow velocity of $\sim$ 550 km s$^{-1}$, we 
identify this higher  velocity with a  possible increase in speed and decrease in density 
of the outflow.  At this 
velocity, the wind would have caught up with the slower expanding ejecta and 
reached the 50 AU photospheric  radius of the  cool dense wind in 
about 120 days, or at about the time the rapid decline began. We suggest that 
this transition corresponds  to when the ejecta becomes optically thin.  The 
faster wind may thus be associated with a second eruption as suggested for SN1994W (Dessart et al. 2009).

\section{Discussion -- Supernova or Impostor?}  
\label{sec:discuss}    %%%  SECTION 8  

SN2011ht has been called both an impostor and a true supernova. With its high visual
luminosity at maximum it is tempting to assume that it was a terminal explosion,
but there are serious contradictions with the standard expectations for supernovae. 
SN2011ht exhibited relatively modest outflow velocities and 
   a total detected energy release that seem more appropriate for a 
   non-supernova giant eruption.  

Integrating over  SN2011ht's multi-wavelength light curve, 
the luminous energy emitted from discovery to the onset of the decline 
was roughly $2.5 \times 10^{49}$ ergs.  The kinetic energy based 
on our mass-loss estimates would be much smaller.  Additional mass 
loss is possible, but at speeds around 
600 km s$^{-1}$, even 10 $M_\odot$, for example,  would carry less than $4 \times 10^{49}$ ergs.  
{\it There is no evidence for substantially higher velocities 
in any of our spectra before the decline,} and later only moderately 
faster values in small amounts of material.
The data therefore strongly suggest $E < 10^{50}$ km s$^{-1}$, 
an order  of magnitude less than a supernova's typical kinetic 
plus luminous output.   The observed amount of energy is present 
in the outer parts of a massive hot star,     
if a suitable instability exists to make it available.

The analysis in {\S}5.1 presents difficulties for a hypothetical 
massive SN blast wave. Three different modes of reasoning all led to a 
characteristic size of 30--60 AU for both the continuum and the 
hydrogen emission lines, during a two-month interval around maximum 
brightness.  The inferred mass in the emission region was then small, 
$M_\mathrm{em} << 1 \; M_\odot$.  An initial blast wave faster than 
1500 km s$^{-1}$ would have passed that radius several weeks earlier. 
In that case, one must explain why the hypothetical massive shock did 
not heat the observed gas and sweep it away.  
If the spectra represent successive inner 
layers following a blast wave outward, then why did the 
observed velocity remain fairly constant?  (A simple model would 
predict a continuously decreasing velocity, resembling a Hubble flow.      
Naively, at least, the hydrogen-rich spectrum also seems a little 
surprising for inner layers of a SN outflow.)  Generally speaking, 
the observed   behavior looked like a continuous, moderate-mass outflow 
from a fixed center, i.e., from a surviving star.  

A true SN may be able to evade the above objections 
  {\it by strongly departing from spherical symmetry.}  Conceivably 
  the main shock wave and kinetic energy were restricted 
  to a range of spatial directions, with slower outflows 
  in other directions.  One can imagine either an axially 
  symmetric explosion (pictorially resembling a GRB and/or 
  $\eta$ Car), or a fully asymmetric event.  With axial  
  symmetry, for instance, our observations might refer to 
  equatorial material while a massive shocked polar outflow 
  was too hot to see.  We cannot develop this 
  possibility here, but it has relevance to Type IIn 
  supernovae in general.

As an anonymous referee comments, an electron-capture supernova has been proposed by 
some authors 
\citep{Bott09,Thompson08} to explain the less luminous outbursts from lower mass stars such as
the ILRTs. Although a wide range of masses are possible for the progenitor based on the 
constraints discussed in Paper I,  there is no observational information that the progenitor 
of SN2011ht was a lower mass star. Its light curve and spectral evolution 
differ from the ILRTs.  The electron-capture 
   idea may be possible, but it is essentially a conjecture, presupposing the event to have been 
 a supernova, and   not  based on any clear implications from the data.   
Arguments for an unfamiliar type of SN based  on the light 
curve are intrinsically weak, because there is no obvious discrepancy 
between the observed shape and a non-SN eruption.  Spectroscopy is more critical.

The presence of X-ray emission identified with SN2011ht in Paper I 
initially appeared  to support 
a supernova interpretation. Although there was no significant X-ray flux measured at early times with
the {\it Swift} X-Ray Telescope, it was reported later. 
This later measurement  however 
is now in doubt. Observations with {\it Chandra} apparently do not confirm that SN2011ht was the X-ray source
(Pooley 2012).  Although, strictly speaking, X-rays inicate a supernova only indirectly. Giant eruptions may also have shocks and X-rays.

In this paper we have emphasized the lack of clear 
evidence for high velocities in SN2011ht's eruption; but even 
if a shock wave occurred, it need not have been a {\it blast wave\/} 
resulting from a central explosion.  When super-Eddington 
radiation drives an eruption like giant outbursts or 
$\eta$ Car,  the material automatically becomes inhomogenous 
(Owocki \& Shaviv 2012).  Material in at least the lower-density  
regions can be accelerated to supersonic speeds.  As 
Davidson (2012) remarked, one consequence might be a shock at the 
beginning of the eruption, which in principle can accelerate 
outward through the decreasing density gradient of pre-eruption 
material.  In order to verify a true core-explosion SN (a 
bona fide shock wave formed near the center), the shock must 
be one or two orders of magnitude more massive and energetic 
than anything seen directly in SN2011ht.   Perhaps it really 
was a terminal explosion, but if so the main SN kinetic energy 
was remarkably inconspicuous.  

The steep declines  observed in the light curves of SN1994W and SN2009kn have been attributed to a  
 very low production of $^{56}$Ni, assuming that they were true supernovae. We observe a similar 
steep decline in SN2011ht which by analogy would imply an equally low 
amount of $^{56}$Ni. Indeed, Dessart et al. used this low result (0.0026 -- 0.015M$_{\odot}$, Sollerman et al. 1998) to also infer that 
SN1994W might not be a terminal explosion. If the event was not really a supernova, then no $^{56}$Ni is expected. Although dust formation was also observed in these 
two objects plus other Type IIn's, in SN2011ht we detected the dust 
relatively early.  The integrated luminosities from the SEDs
at later times  indicate that the bolometric  luminosity was decreasing rapidly and the
eruption was ending. The decrease in the luminosity would
have facilitated the formation of dust. In \S {6} we showed that dust formation might
account for $\sim$ one-fourth of the decline. The decline was thus likely due to 
a combination of 
the cessation of the eruption accompanied by dust formation.  The eruption 
therefore lasted about one  year.  Identifying the post-decline brightness 
level (Figure \ 1) with radioactive 
  decay would constitute another assumption, not a clear deduction, since 
  a giant eruption object can produce some post-event light without 
  radioactivity (e.g., $\eta$ Car did).

Some of the above comments may also relate to other outbursts labeled as 
Type IIn supernovae.  To explain the narrow emission lines and 
their high luminosities, most authors assume that the terminal
eruption collides with a prevously existing circumstellar medium  
ejected in events that occurred 
years before the SN, sometimes many years earlier. 
CSM of that type 
should be located hundreds or thousands of AU from the star.  But 
Thomson-scattering line profiles combined with moderately hot continua 
seem more appropriate for size scales less than 300 AU, and less than 
100 AU in most cases -- less than a year of travel time for the 
outflowing gas.  This casts some doubt on the concept of separate 
events in some of the objects.
Indeed in  several of these examples, especially SN1994W, SN2009kn, etc., the 
previous eruption is 
interestingly proposed to occur only about one year prior to the supernova explosion. 
In Paper I we 
suggested the same model for SN2011ht. If these objects are in fact supernovae, could
 this prior eruption be related to an 
interior instability just before the terminal eruption, or is this an example of the 
pulsational pair instability that may occur in very massive stars? The ``supernova impostors'', or 
giant eruptions,  are spectroscopically very similar to the Type IIn's. The primary distinction 
is the luminosity at maximum (see Van Dyk and Matheson 2012 for a review and references therein). So  these objects could be 
examples of giant eruptions in which the high luminosity is due to interacting 
ejecta from two non-terminal eruptions as suggested by Dessart et al. for SN1994W.  
In the case of SN 2011ht,  the observations are consistent with {\it one  
continuous outflow event} that began early in 2011 and grew rapidly 
after September.  Since there is no clear theoretical model
for giant eruptions, we cannot exclude the possibility of high visual brightnesses.

Motivated by the velocities and observable energy, we suggest  
   that SN2011ht may have been  a giant eruption driven by super-Eddington radiation 
pressure. The fact that it 
   exceeded the Eddington Limit by a large factor is not a fatal 
   objection, because (1) adequate theory has not been developed, and 
   (2) the data strongly suggest that a hyper-Eddington flow did 
   in fact exist for many weeks, independent of the cause 
   (\S {3} and \S {5} above). Our observations 
and its pre-eruption record  suggest the onset of an eruption only about 6 months before the  discovery. SN2011ht  was thus an ongoing high mass loss episode with an 
initial outburst that produced the slower, denser ejecta.  The faster, lower
density wind may be identified with a second outburst or an accelerated wind 
responsible for the post discovery visual and UV brightening. 
As already noted the post-plateau spectrum, dominated by the faster wind or ejecta,  
closely resembles the spectra of several giant eruption objects.

Thus SN2011ht and its twins, SN1994W and SN2009kn, bring into question the nature of the
Type IIn supernovae, or at least some of them. The true measure of a supernova is 
a terminal explosion, while for an impostor, the star survives. The likely 
progenitor of  SN2011ht however was very faint with an upper magnitude limit 
at $g$ = 22.8 mag (Paper I) which together  with the current  
formation of dust, makes it unlikely we will observe the survivor, if it exists, for some time. If SN2011ht was a true supernova, then the eruption must have been quite 
non-spherical to separate the unseen blast wave from the gas that was observed. 

\acknowledgements
It is a pleasure to thank the Directors of the Space Telescope Science Institute and the 
Gemini Observatory for their allotment of Director's Discretionary Time for 
the STIS/MAMA and NIRI observations with {\it HST} and Gemini-North, respectively. 
This paper uses data taken with the MODS spectrographs built with funding from NSF grant AST-9987045 and the NSF Telescope System Instrumentation Program (TSIP), with additional funds from the Ohio Board of Regents and the Ohio State University Office of Research.
R. Humphreys thanks Perry Berlind and the MMT staff for the Hectospec spectra and 
Luc Dessart for useful comments on Thomson scattering profiles. We also thank 
Peter Roming for providing the later data from {\it Swift} and Chris Kochanek for 
useful remarks on an early draft of this paper.  Research on massive stars by 
R. Humphreys and K. Davidson is supported by  the National Science Foundation 
AST-1019394.

{\it Facilities:} \facility{LBT/MODS1, MMT/Hectospec, HST/STIS, LBT/Lucifer, GEMINI/NIRI}

%%%Tables

\input{Table1.tex}

\input{Table2.tex}

\input{Table3.tex}

\input{Table4.tex}

\input{Table5.tex}

\input{Table6.tex}

\input{Table7.tex}

%% Figures

\begin{figure}
\figurenum{1}
\epsscale{1.0}
\plotone{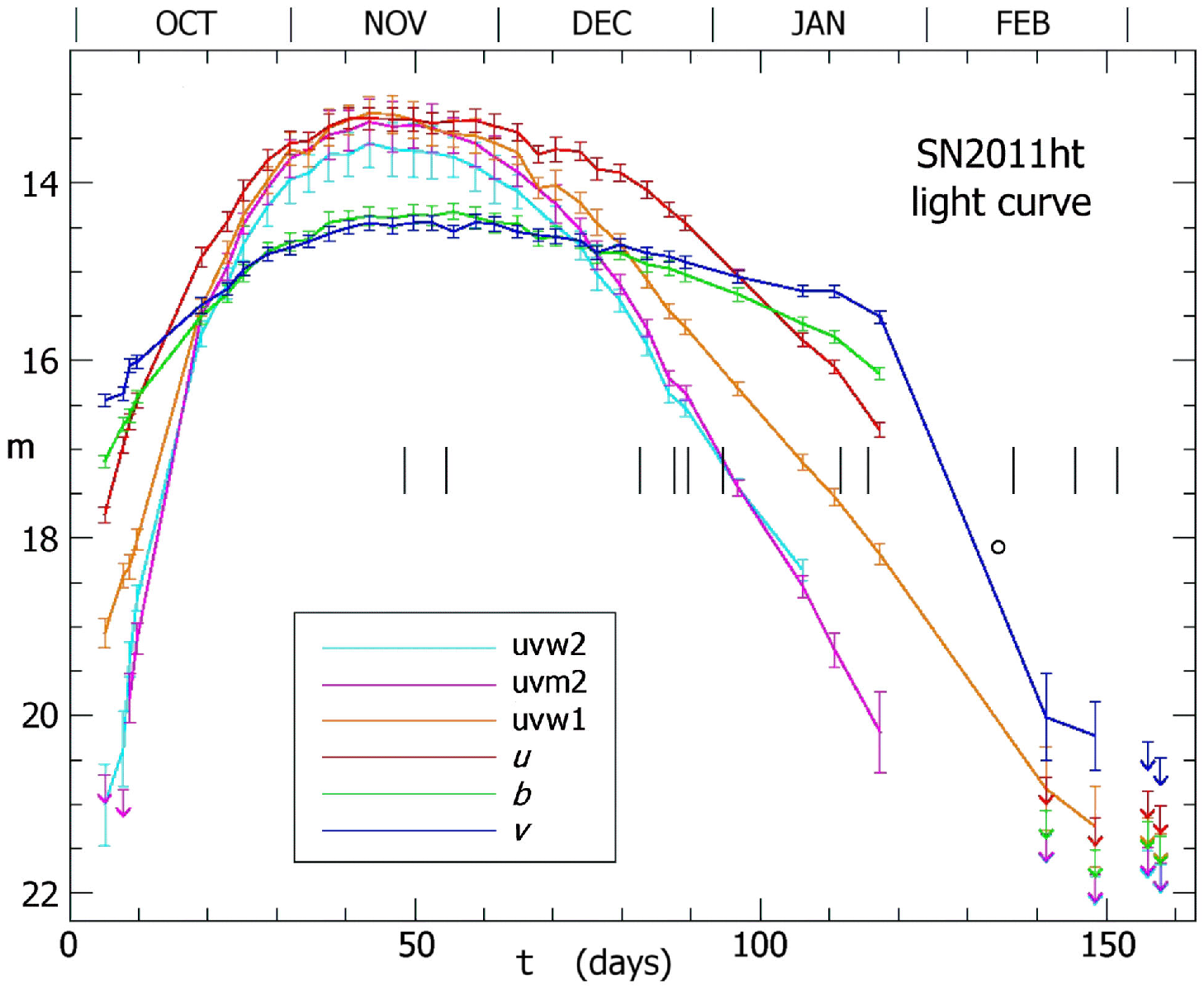}
\caption{The {\it Swift} light curve. The zero-point for $t$ is
MJD 55833 = 2011 September 29. The tic marks show the times of the spectra discussed in this paper. The open circle is an approximate V magnitude from Hurst (2012).}
\end{figure}

\begin{figure}
\figurenum{2}
\epsscale{1.0}
\plotone{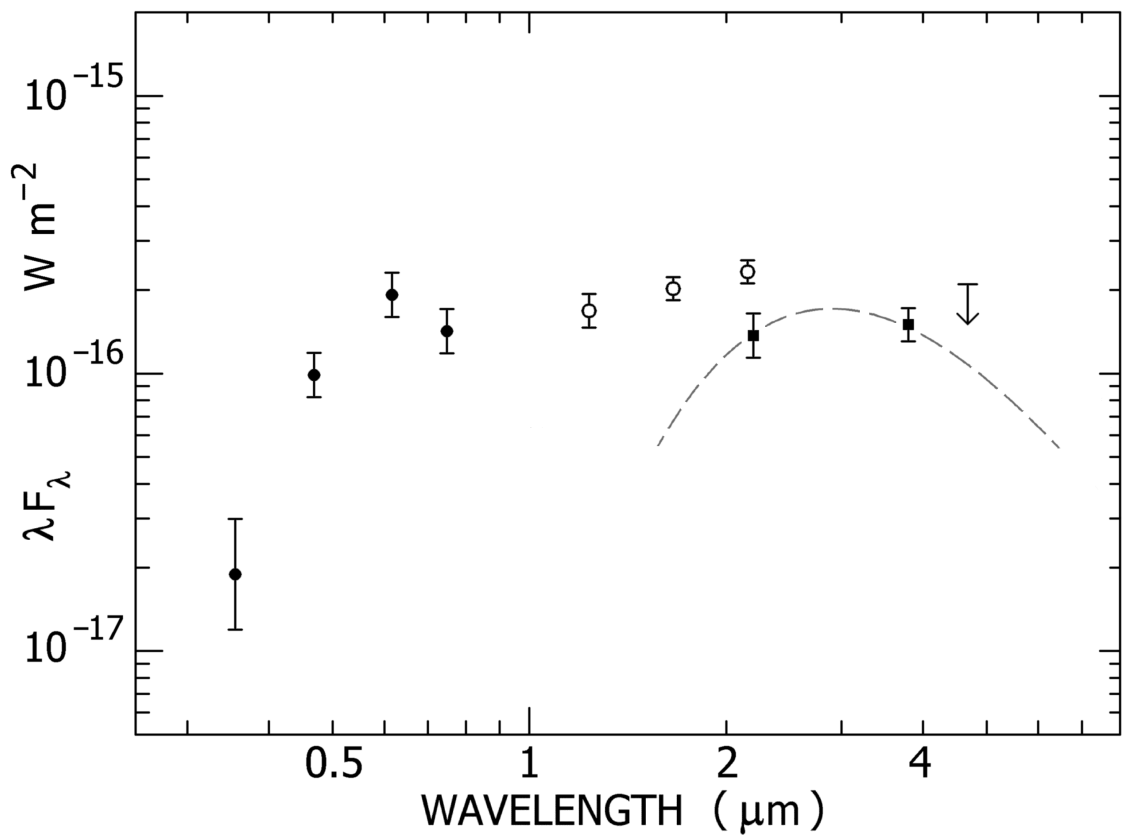}
\caption{The post-plateau spectral energy distribution. The JHK$_{s}$ photometry from 13 March 2012 is shown as open circles, the  $ugri$ magnitudes from 25 March 2012 as filled circles,
  and the KL$^{\prime}$M$^{\prime}$ photometry from 03 May 2012 as filled squares. Even though there is a gap of 7 weeks between the the two K-band measurements and SN2011ht had faded, the L$^{\prime}$ flux and M$^{\prime}$ 
upper limit confirm that SN2011ht formed dust during its rapid decline. The dashed curve
shows the SED produced by dust at 900--1000 K, depending on the detailed emissivity
function.
}
\end{figure}

\begin{figure}
\figurenum{3}
\epsscale{0.8}
\plotone{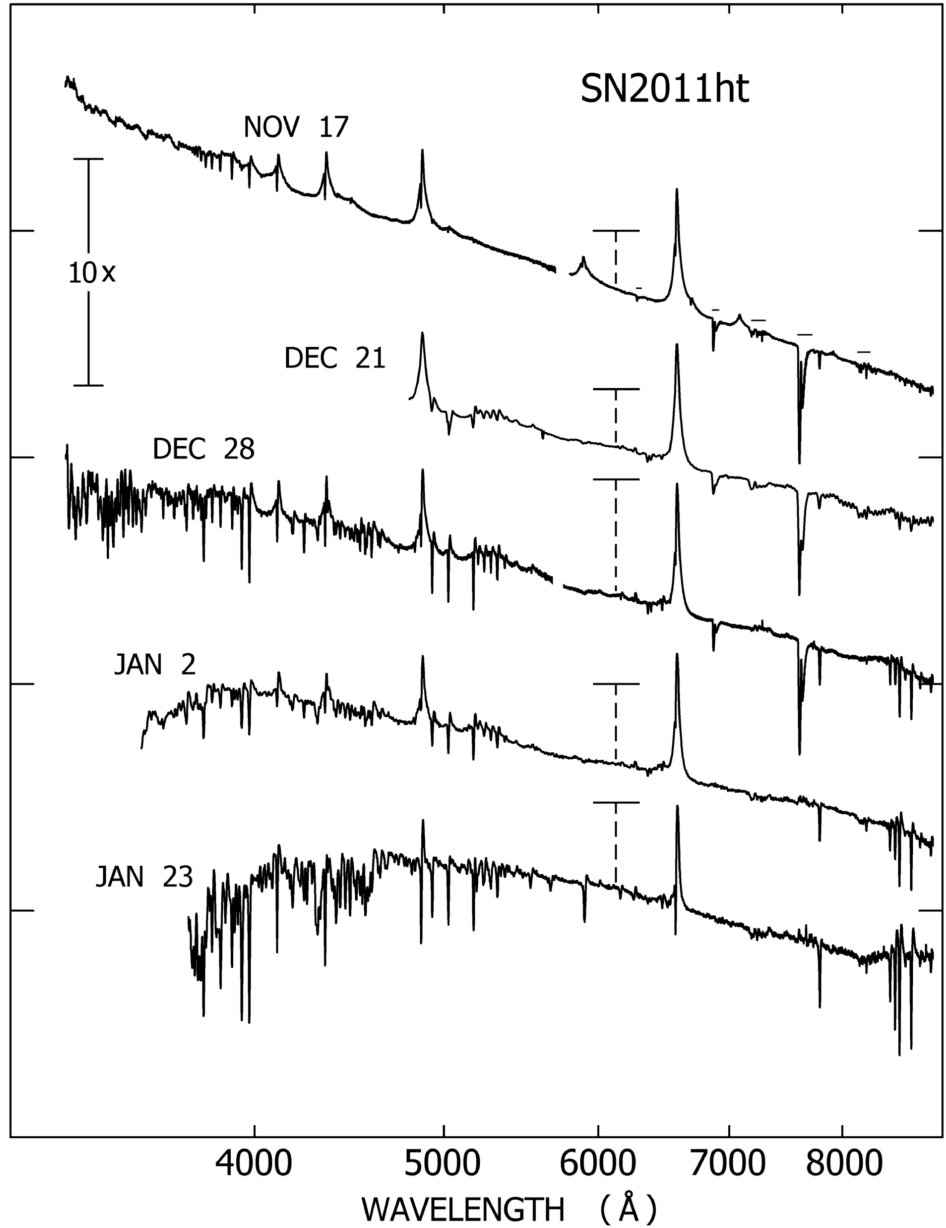}
\caption{The spectra of SN2011ht from 17 November 2011 to 23 January 2012 showing the transition from the hot wind to the cool dense wind. See Table 3 for the details of each spectrum. The 
vertical scale is log F$_{\lambda}$ and the bar above each spectrum marks the 
10$^{-14}$ ergs cm$^{-2}$ s$^{-1}$ {\AA}$^{-1}$ flux level. 
The last spectra from February 2012 are not included here because the flux had 
declined far below the level on 23 January, see Figure 7. The positions of the telluric bands due to O$_{2}$ and H$_{2}$O are marked with a short bar. 
}
\end{figure}

\begin{figure}
\figurenum{4}
\epsscale{1.0}
\plotone{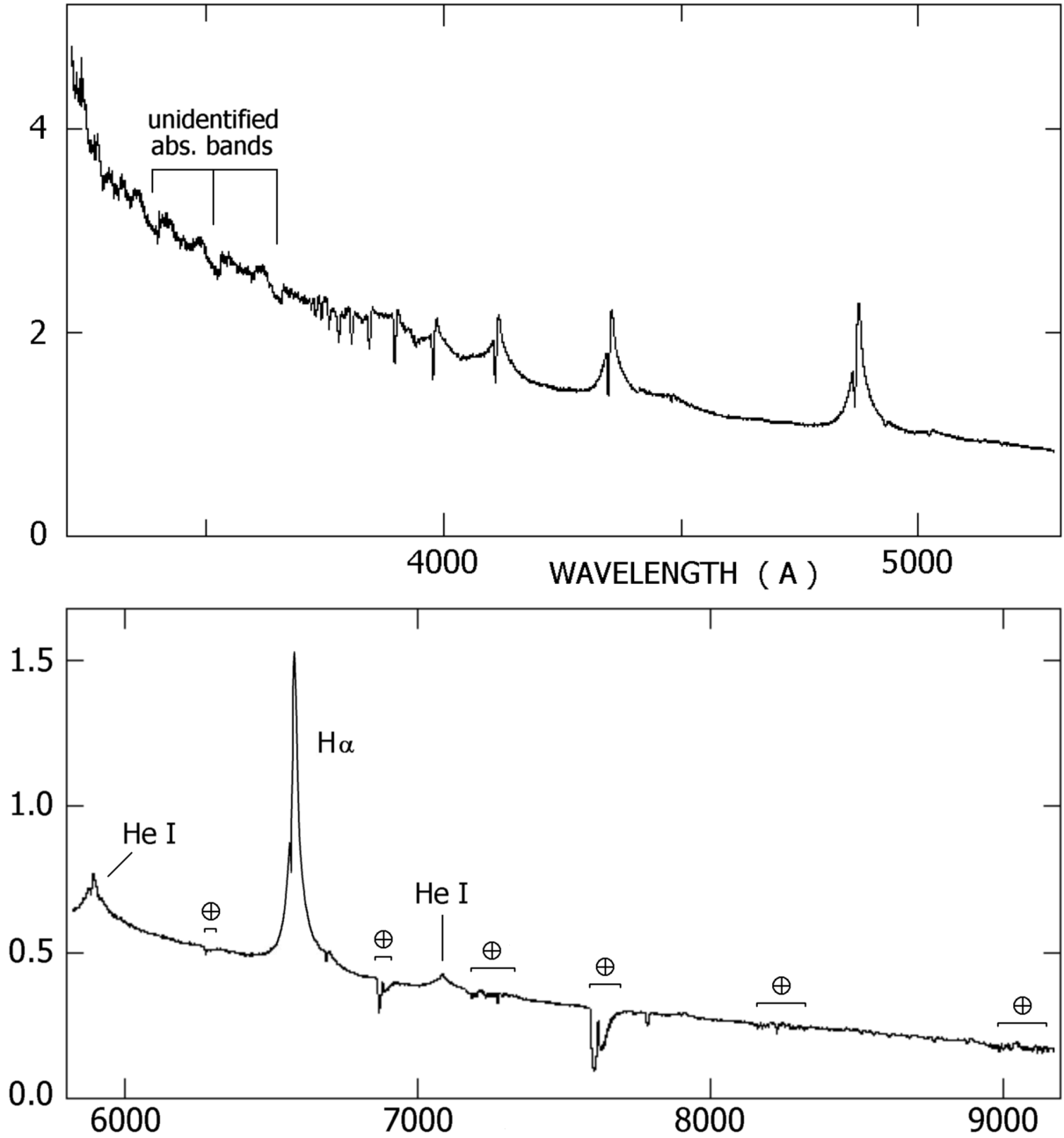}
\caption{The blue (upper) and red (lower) MODS1 spectra from 17 November 2012 during the dense, hot wind stage. The vertical flux scale is in units of 10$^{-14}$ ergs cm$^{-2}$ s$^{-1}$ {\AA}$^{-1}$. The very broad Thomson scattering wings in the Balmer lines and in He I $\lambda$7065 and $\lambda$5876 are obvious. The unidentified bands in the near-UV are marked, and the telluric bands are marked with $\bigoplus$.}
\end{figure}

\begin{figure}
\figurenum{5}
\epsscale{1.0}
\plotone{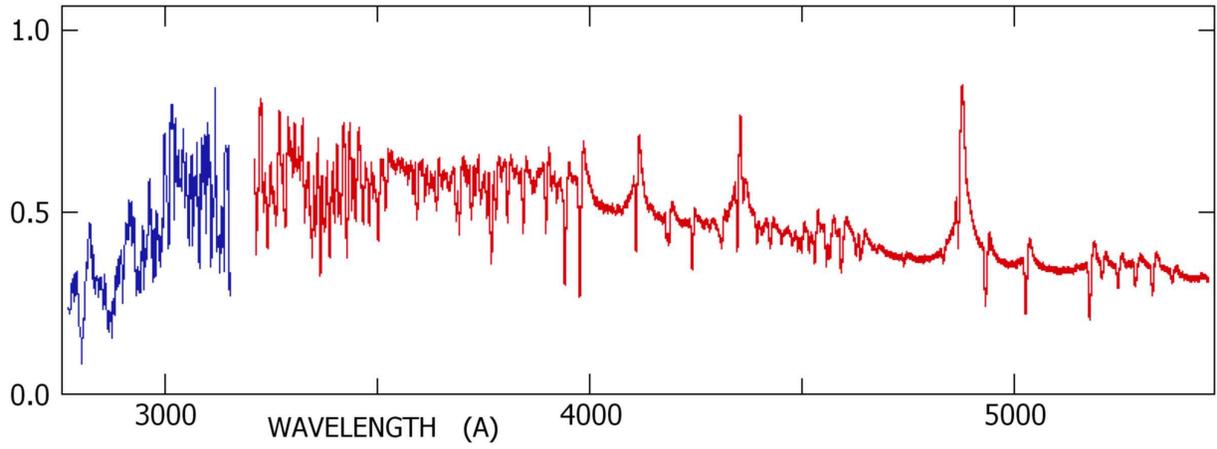}
\caption{The MODS1 spectrum from 26 December 2011 with the nearly overlapping red side of the STIS/MAMA NUV2376 spectrum obtained the same day. The vertical flux scale is in units of 10$^{-14}$ ergs cm$^{-2}$ s$^{-1}$ {\AA}$^{-1}$.  Note the appearance of deep Fe II and Ca II H and K absorption lines.}
\end{figure}

\begin{figure}
\figurenum{6}
\epsscale{1.0}
\plotone{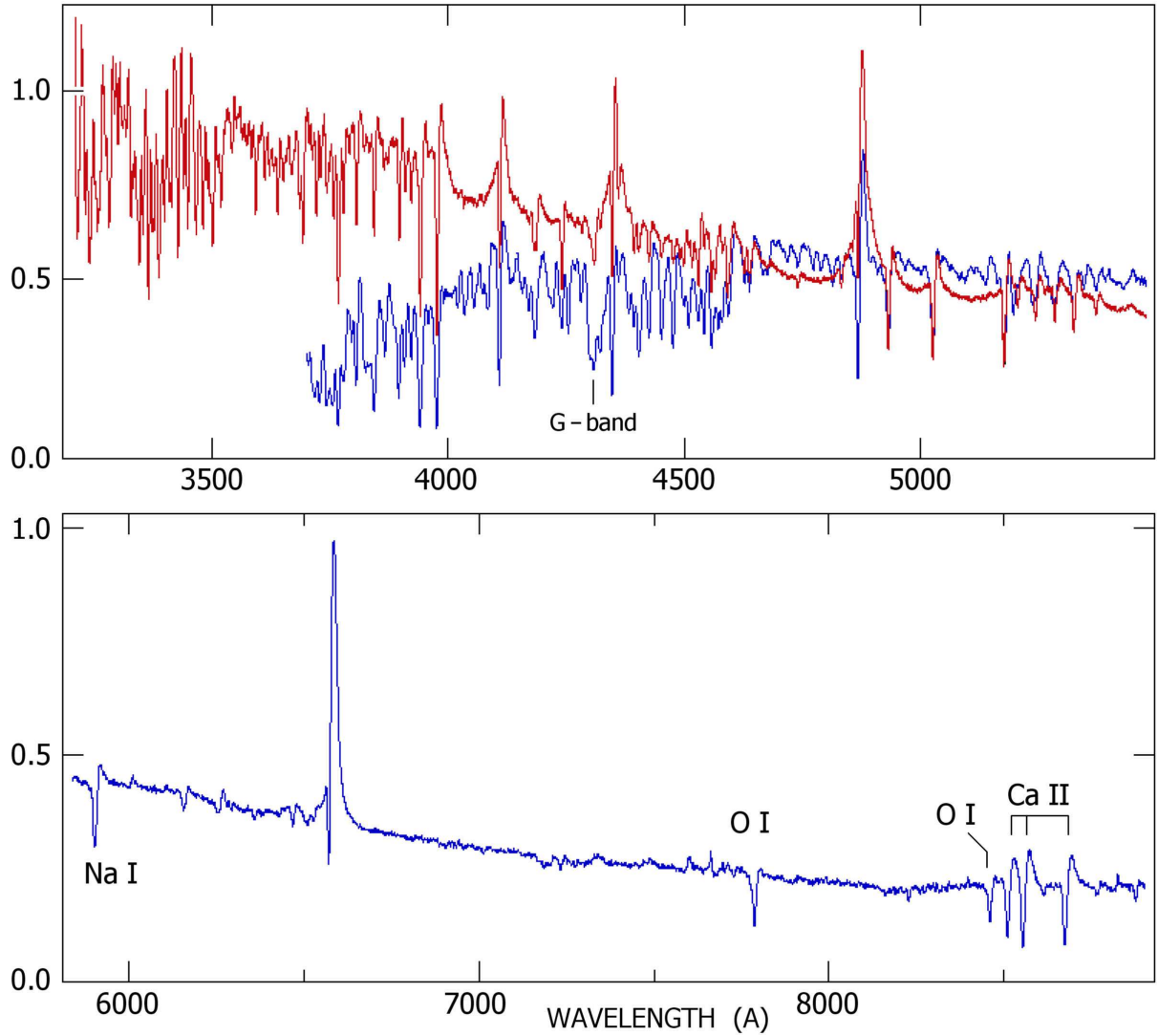}
\caption{Upper: The MODS1 26 December 2011 blue spectrum with the MMT/Hectospec spectrum from 23 January 2012 illustrating the development of the cool dense wind. Lower: The MMT/Hectospec red spectrum from 23 January 2012. The Ca II triplet shows well developed
P Cygni profiles. The vertical flux scale is in units of 10$^{-14}$ ergs cm$^{-2}$ s$^{-1}$ {\AA}$^{-1}$.}
\end{figure}

\begin{figure}
\figurenum{7}
\epsscale{1.0}
\plotone{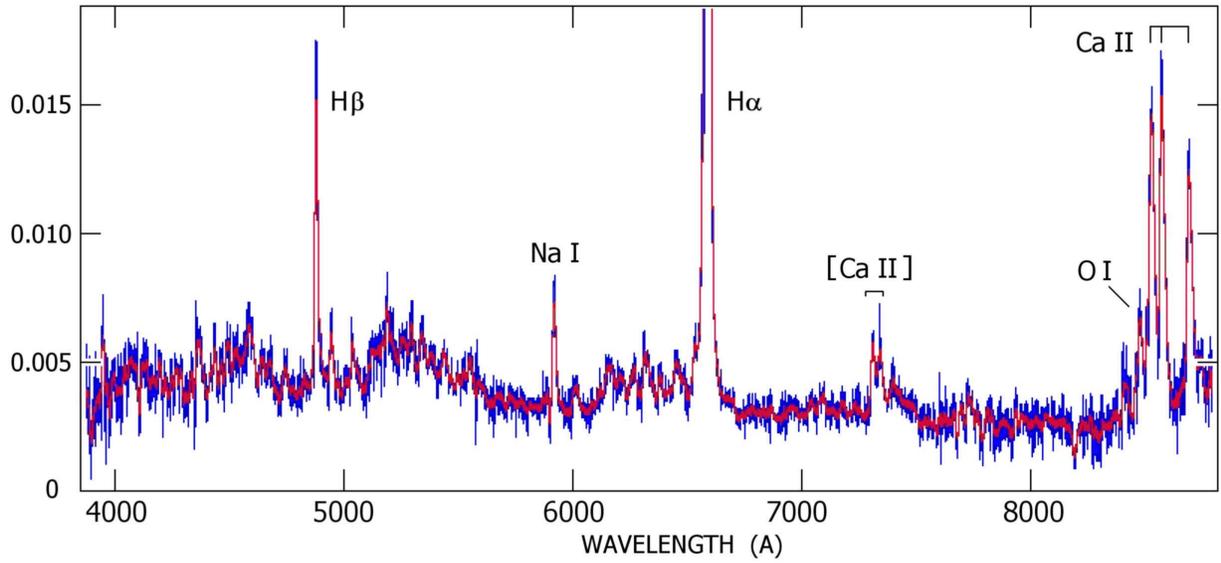}
\caption{Two post-plateau spectra obtained during the photometric decline from 13 February 2012 with the MMT/Hectospec in blue and from 28 February with the ARC DIS in red. The vertical flux scale is in units of 10$^{-14}$ ergs cm$^{-2}$ s$^{-1}$ {\AA}$^{-1}$.}
\end{figure}

\begin{figure}
\figurenum{8}
\epsscale{0.8}
\plotone{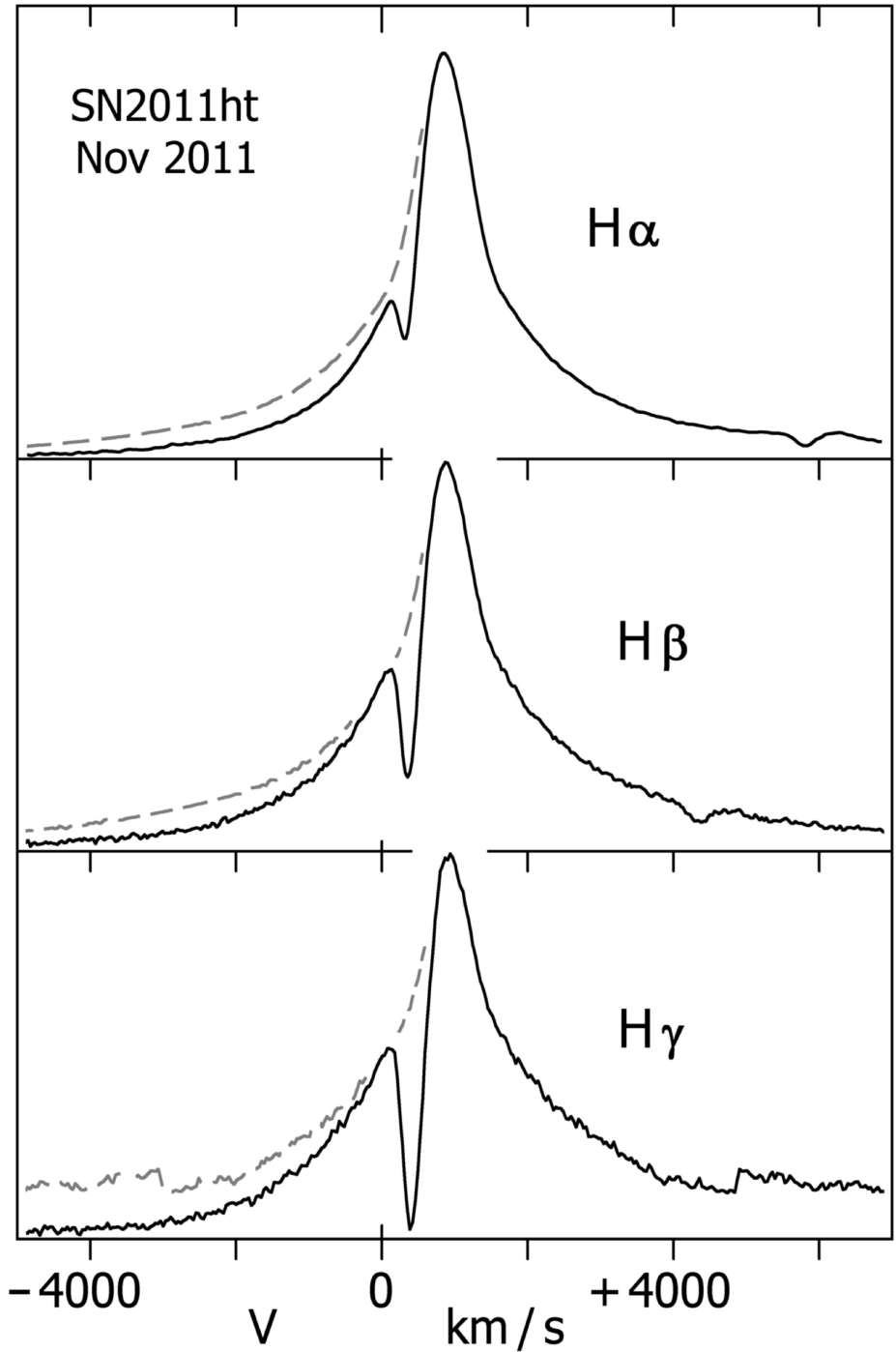}
\caption{The H$\alpha$, H$\beta$ and H$\gamma$ emission profiles from the 17 November 2011 MODS1 spectrum. To illustrate the asymmetry we mirror the red wing as a dashed curve on the left side 
of each profile; compare with Figure 11 in Dessart et al.  Weak He I emission  on the red side of H$\alpha$ and  H$\beta$ is clearly
too weak to alter the basic profile. Heliocentric velocities are shown.}
\end{figure}

%% The following command ends your manuscript. LaTeX will ignore any text
%% that appears after it.

\end{document}

%% file: Table1.tex
\begin{deluxetable}{lccccccc}
\tablewidth{0 pt}
\tabletypesize{\footnotesize}
\tablenum{1} 
\tablecaption{{\it Swift} UVOT Late Times Photometry of SN 2011ht}
\tablehead{
\colhead{Date UT} &
\colhead{Day\tablenotemark{a}} &   
\colhead{} &
\colhead{} &
\colhead{Observed Magnitudes\tablenotemark{b}} &
\colhead{} &
\colhead{} &
\colhead{}  \\ 
\colhead{JD2450000+} &
\colhead{} & 
\colhead{$uvw2$} &
\colhead{$uvm2$} &
\colhead{$uvw1$} &
\colhead{$u$} &
\colhead{$b$} &
\colhead{$v$}  
}
\startdata      
5950.7 & 117 & 19.96(29) &  19.95(**) &  18.18(12)  &  16.79(08) &  16.15(07) &     15.50(07) \\ 
5974.8 & 141 &  21.35(**) &  21.33(**) &  20.82(47)  &  20.69(**) &  21.07(**) &     20.02(49) \\ 
5981.9 &  148 & 21.82(**) &  21.79(**) &  21.25(45)  &  21.15(**) &  21.51(**) &   20.23(39)\\  
5989.5 &  156 & 21.52(**) & 21.49(**)  &  21.15(**)  &  20.84(**) &  21.19(**) &    20.31(**) \\ 
5991.4 &  158 & 21.69(**) & 21.66(**)  &  21.34(**) &   21.02(**) &  21.37(**) &     20.48(**) \\  
\enddata
\tablenotetext{a}{The day number since discovery on 29 September 2011 (Boles et al 2011).} 
\tablenotetext{b}{Values in parenthesis are the errors. 3$\sigma$ upper limits are marked with **.}
\end{deluxetable}

%% file: Table2.tex
\begin{deluxetable}{lcccccccccc}
\rotate
\tablewidth{0 pt}
\tabletypesize{\footnotesize}
\tablenum{2} 
\tablecaption{Late Times Visual and Infrared Magnitudes}
\tablehead{
\colhead{JD} &
\colhead{Day} & 
\colhead{$u$} &
\colhead{$g$} &
\colhead{$r$} &
\colhead{$i$} &
\colhead{$J$} &
\colhead{$H$} & 
\colhead{$K_{s}$} &  
\colhead{$L'$} & 
\colhead{$M'$}
}
\startdata      
5999.6  &  166  &   \nodata  & \nodata  & \nodata  & \nodata &  18.4$\pm$0.15 &  17.4$\pm$0.10 &  16.5$\pm$ 0.10  & \nodata  & \nodata \\
6011.9  &  178  &  23.0 $\pm$ 0.5  &  20.9 $\pm$ 0.2 & 19.9 $\pm$ 0.2 & 20.0 $\pm$ 0.2 & \nodata & \nodata  & \nodata & \nodata  & \nodata \\
6050.2  &  217  &  \nodata  & \nodata  & \nodata  & \nodata &   \nodata  & \nodata & 17.0 $\pm$ 0.2 & 15.3 $\pm$ 0.15 & $>$14.3 \\  
\enddata
\end{deluxetable}

%% file: Table3.tex
\begin{deluxetable}{lllcl}
\tablenum{3}
\tablecaption{Journal of  Spectroscopic Observations}
\tablewidth{0pt}
\tablehead{
\colhead{Instrument} &  \colhead{Date (UT)} & \colhead{Day} &  \colhead{Integration (sec)} & \colhead{Comment/Stage\tablenotemark{a}} 
}
\startdata
LBT/MODS1  &  2011-11-17  &  49   & 900 & Hot wind   \\
LBT/MODS1  &  2011-11-23  &  55   & 900 & Hot wind    \\  
HET/LRS    &  2011-12-21  &  83   & 600 &  Transition    \\
LBT/MODS1  &  2011-12-26  &  88   &  900 & Transition     \\
STIS/MAMA  &  2011-12-26  &  88   &  400 & G230L/NUV2376 \\
STIS/MAMA  &  2011-12-26  &  88   &  400 & G230M/NUV2800 \\
STIS/MAMA  &  2011-12-26  &  88   &  600 & G140L/FUV1425 \\
LBT/MODS1  &  2011-12-28  &  90   &   900 & Transition       \\
ARC/DIS    &  2012-01-02  &  95   &  600    & Transition        \\
ARC/DIS    &  2012-01-19  &  112  &   700    &  Cool wind       \\
MMT/Hectospec & 2012-01-23 & 116   &  900 &  Cool wind   \\
MMT/Hectospec & 2012-02-13 & 137   &   900 &  Decline    \\
ARC/DIS    &    2012-02-22 & 146   &  900    & Decline      \\
ARC/DIS    &    2012-02-28  & 152  &  1200   & Decline       \\
\enddata
\tablenotetext{a}{See \S {4}}
\end{deluxetable}

%% file: Table4.tex
\begin{deluxetable}{llll}
\tabletypesize{\footnotesize}
\tablenum{4} 
\tablecaption{Radial Velocity Summary for SN2011ht - hot wind phase\tablenotemark{a}}
\tablewidth{0pt}
\tablehead{
\colhead{Identification}  &  \colhead{Emission} &  \colhead{Absorption} & \colhead{Difference} \\
           &    \colhead{km s$^{-1}$} &  \colhead{km s$^{-1}$} &  \colhead{km s$^{-1}$}
}
\startdata 
Hydrogen   &     &    &    \\
H$\alpha$  &  926        &     384     &    542   \\
H$\beta$   &  950        &     429     &    521   \\
H$\gamma$  &  995        &     470     &    525   \\
H$\delta$  & 1046        &     490     &    556   \\
H$\epsilon$    & 1096        &     514     &    582   \\
H 8        & 1072        &     532     &    540   \\
mean       &  1014 $\pm$ 62  & \nodata &   544 $\pm$ 20 \\
H 9        & \nodata     &     540     &   \nodata \\
H10        & \nodata     &     569     &   \nodata \\ 
H11        & \nodata     &     565     &   \nodata \\      
H12        & \nodata     &     552     &   \nodata   \\   
H13        & \nodata     &     530     &     \nodata \\   
H14        & \nodata     &     556     &     \nodata \\ 
H15        & \nodata     &     598     &     \nodata \\ 
mean       &  \nodata    &  558 $\pm$ 20 &  \nodata \\  
         &             &                      &        \\
Helium   &         &       &      \\
$\lambda$7065 & 955      &    \nodata    &   \nodata     \\
$\lambda$6678 & 1033     &   615         &   418     \\
$\lambda$5876 & 1041     &   567         &   474      \\
$\lambda$5016 &  1116    & 646          &    470     \\
$\lambda$4922 &  1082  & 695          &   387     \\
$\lambda$4471 & 976       &  610     &  366     \\ 
$\lambda$4026 & \nodata  &  633          &  \nodata     \\
$\lambda$3819 &  \nodata  & 597          &  \nodata     \\   
mean           &  1033 $\pm$ 56 &  615 $\pm$ 17 & 423$\pm$ 43    \\
         &             &                      &        \\  
Other   &         &       &      \\
Ca II K  &     \nodata &      564        &   \nodata    \\ 
O I $\lambda$7774 &    \nodata &       529        &   \nodata    \\ 
\enddata
\tablenotetext{a}{Line indentifications and Doppler velocities measured from the MODS1 spectrum from 17 Nov. 2011.}
\end{deluxetable}

%% file: Table5.tex
\begin{deluxetable}{llll}
\tabletypesize{\footnotesize}
\tablenum{5} 
\tablecaption{Radial Velocity Summary for SN2011ht Post-Maximum Spectra}
\tablewidth{0pt}
\tablehead{
\colhead{Identification}  &  \colhead{Emission} &  \colhead{Absorption} & \colhead{Difference} \\
           &    \colhead{km s$^{-1}$} &  \colhead{km s$^{-1}$} &  \colhead{km s$^{-1}$}
}
\startdata 
{\it Transition}\tablenotemark{a} &           &       & \\   
Hydrogen   &      &      &     \\
H$\alpha$  &   928       &   338       &   590     \\
H$\beta$   &   962       &   370       &   592     \\
H$\gamma$  &   947       &   435       &   512     \\
H$\delta$  &   1031      &   468       &   563     \\
H$\epsilon$&   1073      &   \nodata   &   \nodata \\
H 8        &   1072      &   509       &   563     \\
H 9        &   1103      &   548       &   555     \\
mean       &   1017 $\pm$ 65  & \nodata   &  563 $\pm$ 26 \\
H 10       &   \nodata   &   561       &  \nodata \\
H 11       &   \nodata   &   557        &  \nodata \\ 
H 12       &   \nodata   &   528        & \nodata \\
H 13       &   \nodata   &   595        & \nodata \\ 
mean       & \nodata   &   560 $\pm$ 24  & \nodata \\
           &             &                      &        \\ 
Other         &             &                      &        \\
Ca II K  &     \nodata &   511         & \nodata \\
Ca II H  &     \nodata &   522         & \nodata \\
Fe II    &  1075 $\pm$ 61 (8) & 522 $\pm$ 31 (8) & 593  $\pm$ 62 (7) \\
O I $\lambda$7774 &  \nodata & 536     &  \nodata \\ 
O I $\lambda$8446  &  \nodata & 529       &  \nodata \\ 
Ca II $\lambda$8498 & \nodata  & 593     &  \nodata   \\
Ca II $\lambda$8542 &   1117    &  527   &  590   \\ 
Ca II $\lambda$8662 &  1036  &  554      &  482          \\ 
mean   & 1076 $\pm$ 57   &  528 $\pm$ 32   & 580 $\pm$ 65 \\  
         &             &                      &        \\
{\it Cool Wind}\tablenotemark{b}      &             &            &    \\   
Hydrogen   &  \nodata    &   \nodata   &   \nodata  \\
H$\alpha$  &  992        &     329     &    663   \\
H$\beta$   &  1086        &    426     &    660   \\
H$\gamma$  &  1057        &    470      &   587   \\
H$\delta$  &  1156        &    490     &    666   \\
mean       &  1073 $\pm$ 59  & \nodata &  644 $\pm$ 33 \\ 
Other         &             &                      &        \\  
Ca II K  &     \nodata &       473        &   \nodata    \\ 
Ca II H  &    \nodata &       555         &   \nodata    \\  
Fe II    &   1104 $\pm$ 22  (3 lines) & 554 $\pm$ 25 (4 lines) & 589 (2 lines) \\ 
Na I D   &   1158        &   445      &  713       \\
O I $\lambda$7774 &    \nodata &       525        &   \nodata    \\ 
O I $\lambda$8446 &    \nodata &       494        &   \nodata    \\ 
Ca II $\lambda$8498 &  1144    &       470        &   674     \\
Ca II $\lambda$8542 &  1113    &       436        &   677    \\
Ca II $\lambda$8662 &  1091    &       436        &   655     \\ 
mean      &   1116 $\pm$ 23   &    533 $\pm$ 35   &   637 $\pm$ 40 \\ 
         &             &                      &        \\  
{\it Warm Thin Wind}\tablenotemark{c}    &           &                &   \\   
Hydrogen   &  \nodata    &   \nodata   &   \nodata \\
H$\alpha$  &  1070       &   315       &  755 \\ 
H$\beta$   &  1123       &   432       & 691  \\ 
H$\gamma$  &  1182       &   470       & 712  \\
mean       &  1125$\pm$ 46 & \nodata  & 719$\pm$ 26  \\
H$\delta$  &  \nodata    &   527       & \nodata \\
H$\epsilon$ & \nodata    &   540       & \nodata \\
Other         &             &                      &        \\
Na I D     &  1285       &  430        &  855   \\
Fe II      &  1158$\pm$ 34 (4 lines) &  \nodata   &   \nodata \\
{[Ca II]} $\lambda$7291  &  732  &  \nodata   &   \nodata  \\ 
{[Ca II]} $\lambda$7324  &  772  &  \nodata   &   \nodata  \\ 
O I $\lambda$8446 &  959       &  \nodata   &   \nodata  \\
Ca II $\lambda$8498 & 971      &  \nodata   &   \nodata  \\
Ca II $\lambda$8542 &  962     &  \nodata   &   \nodata  \\
Ca II $\lambda$8662 &  935     &  \nodata   &   \nodata  \\
\enddata
\tablenotetext{a}{Line indentifications and Doppler velocities measured from the MODS1 spectrum from 26 Dec. 2011.}
\tablenotetext{b}{Doppler velocities are the average from the ARC and MMT/Hectospec spectra from 19 Jan. and 23 Jan.2012.}
\tablenotetext{c}{Doppler velocities are the average from the ARC and MMT/Hectospec spectra from 13 Feb., 22 Feb., and 28 Feb. 2012.}
\end{deluxetable}

%% file: Table6.tex
\begin{deluxetable}{llllll}
\tabletypesize{\footnotesize}
\tablenum{6} 
\tablecaption{Emission and Absorption Velocity Summary}
\tablewidth{0pt}
\tablehead{
\colhead{Stage} &  \colhead{Balmer Em.} & \colhead{He I Em.} & \colhead{Other Em.} & \colhead{P Cyg Abs Difference}  & \colhead{Absorption} \\
                &   \colhead{km s$^{-1}$} &  \colhead{km s$^{-1}$} &  \colhead{km s$^{-1}$}
		&   \colhead{km s$^{-1}$} &  \colhead{km s$^{-1}$}
}
\startdata
Hot Wind   &    1014$\pm$62 & 1033$\pm$56 & \nodata & 544$\pm$20 &  556$\pm$20 \\
Transition &    1017$\pm$65 &  \nodata   & 1076$\pm$57 & 580$\pm$65 & 528$\pm$33\\
Cool Wind  &    1073$\pm$59 & \nodata   &  1116$\pm$23 & 640$\pm$37 & 533$\pm$35\\
Thin Warm Wind & 1125$\pm$46 &  \nodata   & 1172$\pm$28\tablenotemark{a}  &  719$\pm$26 & 534$\pm$6 \\
\enddata
\tablenotetext{a}{Not including Ca II and [Ca II]. See Table 5 and \S {5}.}
\end{deluxetable}

%% file: Table7.tex
\begin{deluxetable}{lllll}
\tabletypesize{\footnotesize}
\tablenum{7} 
\tablecaption{The Thomson Scattering Wings}  
\tablewidth{0pt}
\tablehead{
\colhead{Identification}  &  \colhead{Blue Wing} & \colhead{Blue Wing} & \colhead{Red Wing} & \colhead{Red Wing}\\
                          &  \colhead{$\Delta$$\Lambda$({\AA})}  & \colhead{km s$^{-1}$} & \colhead{$\Delta$$\Lambda$({\AA})}   & \colhead{km s$^{-1}$}
}
\startdata
{\it Hot Wind} (17 and 23 November 2011)  &     &    &    &    \\
H$\alpha$    & -105  & -4726   &  +174    &  7939  \\
He I $\lambda$7065  & -61    & -2584   & +86    &  3636  \\
He I $\lambda$5876  & -58    & -2936   & +108     &   5490  \\
               &     &         &      &           \\
{\it Transition} (26 and 28 December 2012)  &     &    &    &    \\
H$\alpha$    & -55   &  -2518  & +102  &  4637   \\
               &     &         &      &           \\
{\it Cool Wind}  (19 and  23 January  2012) &     &    &    &    \\
H$\alpha$    &  -56     &  -2549  & +73     &  3319 \\
Ca II $\lambda$8542 &  \nodata   & \nodata   & +28      &  987 \\
Ca II $\lambda$8662 &  \nodata   & \nodata   & +26      &  903  \\
\enddata
\end{deluxetable}

%% file: SN2011ht.bbl
\begin{thebibliography}{}

\bibitem[Andrews et al.(2010)]{And}Andrews, J. E., et al. 2010, \apj, 715, 541 
\bibitem[Berger et al.(2009)]{Berger09}Berger, E. et al. 2009, \apj, 699, 1850  
\bibitem[Boles(2011)]{Boles}Boles, T. 2011, CBET 2851
\bibitem[Bond et al.(2009)]{Bond09}Bond, H. E., Bedin, L. R., Bonanos, A. Z., Humphreys, R. M., Monard, L. A. G. B., Prieto, J. L., \&  Walter, F. M. 2009, \apjl, 695, L154 
\bibitem[Botticella et al.(2009)]{Bott09}Botticella, M. T., et al. 2009, \mnras, 
398, 1041
\bibitem[Chugai(2001)]{Chugai01}Chugai, N. N. 2001, \mnras, 326, 1448 
\bibitem[Chugai et al.(2004)]{Chugai04}Chugai,  N. N. et al. 2004, \mnras, 352, 1213 
\bibitem[Cox et al.(1995)]{Cox}Cox, P., Mezger, P.G., Sievers, A., Najarro, F., Bronfman, L., Kreysa, E., \& Haslam, G. 1995, \aap, 297, 168
\bibitem[Davidson(1987)]{Davidson87}Davidson, K. 1987, \apj, 317, 760 
\bibitem[Davidson et al.(1995)]{KD95}Davidson, K., Ebbets, D., Weigelt, G., Humphreys, R. M., Hajian, A. R., Walborn, N. R. \&  Rosa, M. 1995, \aj, 109, 1784  
\bibitem[Davidson and Humphreys(1997)]{DH97} Davidson, K., \& Humphreys, R.M. 1997, Ann.\ Revs.\ Astr.\ Astrophysics 35, 1 
\bibitem[Davidson(2012)]{KD12} Davidson, K. 2012, in Eta Carinae and the Supernova Impostors, Astrophys.\ \& Sp.\ Sci.\ Library 384 (ed.\ K.\ Davidson \& R.M.\ Humphreys, Springer Media, New York), 43
\bibitem[Dessart et al.(2009)]{Dess08}Dessart, L., Hillier, D. J., Gezari, S., Basa, S. \& Matheson, T. 2009, \mnras, 394, 21  
\bibitem[Draine(2011)]{Draine} Draine, B.T. 2011, {\it Physics of the Interstellar and 
Intergalactic Medium} (Princeton University Press)
\bibitem[Foley et al.(2011)]{Foley}Foley, R. J., et al. 2011, \apj, 732, 32  
\bibitem[Fox  et al.(2011)]{Fox}Fox, et al. 2011, \apj, 741, 7 
\bibitem[Hillier, et al.(2001)]{Hillier}Hillier, D. J., Davidson, K., Ishibashi, K. \& Gull, T. 2001, \apj, 553, 837 
\bibitem[Humphreys and Davidson(1994)]{HD94}Humphreys, R.M.\& Davidson, K. 1994, \pasp, 106, 1025
\bibitem[Humphreys et al.(2011)]{RMH11}Humphreys, R.M., Bond, H. E., Bedin, L. R., 
Bonanos, A. Z., Davidson, K., Monard, L.A.G. B.,  Prieto, J. L. \& Walter, F. M. 2011, \apj, 743, 118   
\bibitem[Humphreys(2012)]{RMH12}Humphreys, R.M. 2012, The Astronomer's Telegram, 3895
\bibitem[Hurst(2012)]{Hurst}Hurst, G. 2012, private communication
\bibitem[Jester et al.(2005)]{Jest05}Jester, S., et al. 2005, \aj, 830, 173 
\bibitem[Kankare, et al.(2012)]{Kankare}Kankare, E. et al. 2012, \mnras 
\bibitem[Karaali, Blir \& Tuncel(2005)]{Kar05}Karaali, S., Bilir, S., \& Tuncel, S, 2005, Publ. Astron. Soc. Aust., 22, 24  
\bibitem[Kiewe et al.(2012)]{Kiewe}Kiewe, M. et al. 2012, \apj, 744, 10  
\bibitem[Kochanek(2012)]{csk}Kochanek, C. 2012, private communication 
\bibitem[Martin et al.(2006)]{JCM}  Martin, J.C., Davidson, K., \& Koppelman, M.D. 2006, \aj, 139, 2056
\bibitem[Mehner et al.(2010)]{AM2010}Mehner, A., Davidson, K., Humphreys, R.M., Martin, J.C., Ishibashi, K., Ferland, G.J., \& Walborn, N.R. 2010, \apj, 717, L22 
\bibitem[Mehner et al.(2012)]{AM2012}Mehner, A., Davidson, K., Humphreys, R.M., Ishibashi, K., \& Martin, J.C.  2012, \apj, 751:73
\bibitem[Ofek et al.(2010)]{Ofek}Ofek, E. O., et al. 2010, \apj, 724, 1396 
\bibitem[Osterbrock and Ferland(2006)]{OF}Osterbrock, D. F. \& Ferland, G. F. 2006, {\it Astrophysics of Gaseous Nebulae and Active Galactic Nuclei} {University Science Books} 
\bibitem[Otsuka M., et al.(2012)]{OTsuka}Otsuka, M. et al. 2012, \apj, 744, 26 
\bibitem[Owocki and Shaviv(2012)]{Owocki}Owocki, S., \& Shaviv, N. 2012, in Eta Carinae 
and the Supernova Impostors, Astrophys.\ \& Sp.\ Sci.\ Library 384 (ed.\ K.\ Davidson \& R.M.\ Humphreys, Springer Media, New York), 275
\bibitem[Pastorello, A., et al.(2011)]{Past11}Pastorello, A., Stanishev, V., Smartt, S. J., \&  Fraser, M. 2011, CBET 2851
\bibitem[Pooley(2012)]{Pool}Pooley, D. 2012, The Astronomer's Telegram, 4062
\bibitem[Prieto, J. L. et al.(2011a)]{Prieto11a}Prieto, J. L., et al. 2011, The Astronomer's Telegram, 3749  
\bibitem[Prieto, J. L. et al.(2011b)]{Prieto11b}Prieto, J. L., McMillan, R., \& Bakos, G. 2011, CBET 2903
\bibitem[Roming et al.(2011)]{Rom11}Roming, P. W. A., Pritchard, T.A., \& Brown, P. 2011, ATel 3690
\bibitem[Roming et al.(2012)]{Rom12}Roming, P. W. A. et al. 2012, \apj, 751:92 (Paper I) 
\bibitem[Smith et al.(2009)]{NS09}Smith, N., et al. 2009, \apjl, 697, 49
\bibitem[Smith et al.(2010)]{NS10}Smith, N., et al. 2010, \aj, 139, 1451 
\bibitem[Sollerman et al.(1998)]{Soll}Sollerman, J., Cuming, R. J., \& Lundquist, P. 1998, \apj, 493, 933 
\bibitem[Thompson et al.(2009)]{Thompson08}Thompson, T., Prieto, J. L., Stanek, K
. Z., Kistler, M. D., Beacom, J. F. \&  Kochanek, C. S. 2009, \apj, 705, 1364 
\bibitem[Van Dyk(2012)]{vandyk}Van Dyk, S. D. and Matheson, T. 2012, in Eta Carinae  
and the Supernova Impostors, Astrophys.\ \& Sp.\ Sci.\ Library 384 (ed.\ K.\ Davidson \& R.M.\ Humphreys, Springer Media, New York),  249 
\end{thebibliography}
